\documentclass[12pt]{iopart}

\usepackage[utf8]{inputenc}
\usepackage{amssymb}
\expandafter\let\csname equation*\endcsname\relax
\expandafter\let\csname endequation*\endcsname\relax\usepackage{iopams}  
\usepackage{hyperref}
\usepackage{graphicx}
\usepackage{cite}
\usepackage{hyperref}
\usepackage[]{slashed}
\usepackage[]{bm}     
\usepackage{physics}
\usepackage{lipsum}
\usepackage{dsfont}
\usepackage[dvipsnames]{xcolor}
\usepackage{soul}
\definecolor{bubblegum}{rgb}{0.99, 0.76, 0.8}

\DeclareMathOperator\arccosh{arccosh}

\usepackage{graphicx}
\usepackage{epstopdf}
\usepackage{bbm,graphicx,amsthm}

\usepackage[force]{feynmp-auto}

\def\triangleboxleft{\scalebox{.9}{$\triangleleft$}\kern-.1em\Box}
\def\triangleboxright{\Box\kern-.1em\scalebox{.9}{$\triangleright$}}
\def\dBox{\Box\kern-.1em\Box}
\def\deltabar{{\hat\delta}}

\usepackage{float}
\usepackage[T1]{fontenc} 

\expandafter\let\csname equation*\endcsname\relax
\expandafter\let\csname endequation*\endcsname\relax
\usepackage{amsmath}

\usepackage{etoolbox}

\makeatletter
\def\@mkboth#1#2{}
\newlength\appendixwidth
\preto\appendix{\addtocontents{toc}{\protect\patchl@section}}
\newcommand{\patchl@section}{%
  \settowidth{\appendixwidth}{\textbf{Appendix }}%
  \addtolength{\appendixwidth}{1.5em}%
  \patchcmd{\l@section}{1.5em}{\appendixwidth}{}{\ddt}%
}
\makeatother

\makeatletter
\newcommand{\mainmatter}{%
  \setcounter{footnote}{0}%
  \patchcmd{\@makefntext}{\fnsymbol}{\arabic}{}{}%
  \patchcmd{\@thefnmark}{\fnsymbol}{\arabic}{}{}%
  \def\@makefnmark{\textsuperscript{\arabic{footnote}}}%
}
\makeatother

\hypersetup{colorlinks=true,linkcolor=magenta,anchorcolor=green,citecolor=cyan,filecolor=black,menucolor=black,urlcolor=black}

\begin{document}
\bibliographystyle{iopart-num}
\newcommand{\eprint}[2][]{\href{https://arxiv.org/abs/#2}{\tt{#2}}}

\begin{flushright}
IPhT-t22/03,\	SAGEX-22-14
\end{flushright}

\title[Post-Minkowskian from Scattering
Amplitudes]{The SAGEX Review on Scattering Amplitudes,\\
Chapter 13:  Post-Minkowskian expansion from Scattering
  Amplitudes }

\author{N.E.J. Bjerrum-Bohr$^{1}$, P.H. Damgaard$^{1}$, L. Plant\'e$^{1}$, P. Vanhove$^{2}$}

\address{$1$ Niels Bohr International Academy, Niels Bohr Institute, University of Copenhagen, Blegdamsvej 17, DK-2100 Copenhagen, Denmark}
\address{$2$ Institut de Physique Th\'eorique, Universit\'e Paris-Saclay, CEA, CNRS, F-91191 Gif-sur-Yvette Cedex, France}

\ead{bjbohr@nbi.dk, phdamg@nbi.dk,ludo.plan@hotmail.fr, pierre.vanhove@ipht.fr}
\vspace{10pt}
\begin{indented}
\item[] \today
\end{indented}

\begin{abstract}
{The post-Minkowskian expansion of Einstein's general theory of relativity has received much attention in recent years due to the possibility of harnessing the computational power 
of modern amplitude calculations in such a classical context. In this brief review, we focus on the post-Minkowskian expansion as applied to the two-body problem
in general relativity without spin, and we describe how relativistic quantum field theory can be used to greatly simplify analytical calculations based on the Einstein-Hilbert action.
Subtleties related to the extraction of classical physics from such quantum mechanical calculations highlight the care which must be taken when both positive and
negative powers of Planck's constant are at play. In the process of obtaining classical results in both Einstein gravity and supergravity, one learns new aspects of quantum
field theory that are obscured when using units in which Planck's
constant is set to unity. The scattering amplitude
  approach provides a self-contained framework for deriving the two-body scattering valid in
  all regimes of energy. There is hope that the full impact of amplitude computations in this field
may significantly alter the way in which gravitational wave predictions will advance in the coming years.}
\end{abstract}
\maketitle
\tableofcontents
\mainmatter
\section{Introduction} 
%
The observation of gravitational waves radiated by binary systems of massive astrophysical objects has
opened a new and exciting astrophysical avenue for testing Einstein's
theory of gravity. To unlock the full discovery potential of
gravitational-wave astrophysics, and to keep abreast with modern
observational advancements, the development of new analytical methods
is now urgently needed. This prompts for both refinements and
complements to the existing theoretical framework as well as a general
call for new and more efficient methods of computation. Based on the
evident advantages of relativistic quantum field theory, it has been
suggested to focus on the latter by means of an application of modern
amplitude techniques to the post-Minkowskian expansion of general
relativity~\cite{Damour:2016gwp,Damour:2017zjx,Bjerrum-Bohr:2018xdl,Cheung:2018wkq,Cristofoli:2019neg}. During
the inspiral phase the gravitational field is weak and  perturbation
theory can reliably be applied up to a few cycles before the merging phase.
Progress has been swift. For the case of non-spinning black holes the
relativistic amplitude analysis has run from the second
post-Minkowskian order in the above references to third
post-Minkowskian
order~\cite{Bern:2019nnu,Antonelli:2019ytb,Bern:2019crd,Parra-Martinez:2020dzs,DiVecchia:2020ymx,Damour:2020tta,DiVecchia:2021ndb,DiVecchia:2021bdo,Herrmann:2021tct,Bjerrum-Bohr:2021vuf,Bjerrum-Bohr:2021din,Damgaard:2021ipf,Brandhuber:2021eyq}. Most
recently, the relativistic amplitude approach to general relativity
has even reached fourth post-Minkowskian
order~\cite{Bern:2021dqo,Bern:2021yeh} and, in the probe limit, all
the way up to fifth post-Minkowskian
order~\cite{Bjerrum-Bohr:2021wwt}. In this brief review we will
attempt to describe how this rapid sequence of events unfolded. The
corresponding case of spinning black holes is evidently of great
phenomenological importance but the corresponding description in terms
of the amplitude approach to general relativity is far more complex
compared to the non-spinning case. We shall not be able to cover that 
fascinating story of spin but refer to some relevant papers
here~\cite{Guevara:2017csg,Vines:2017hyw,Arkani-Hamed:2017jhn,Guevara:2018wpp,Vines:2018gqi,Chung:2018kqs,Guevara:2019fsj,Maybee:2019jus,Arkani-Hamed:2019ymq,Damgaard:2019lfh,Aoude:2020onz,Chung:2020rrz,Bern:2020buy,Haddad:2020tvs,Guevara:2020xjx,Kosmopoulos:2021zoq,Bautista:2021wfy,Haddad:2021znf,Chen:2021qkk,Jakobsen:2022fcj}. There
is also a parallel development to the post-Minkowskian expansion based
on the world-line
approach~\cite{Kalin:2020fhe,Mogull:2020sak,Jakobsen:2021smu,Jakobsen:2021lvp,Jakobsen:2021zvh,Dlapa:2021npj,Jakobsen:2022fcj}
that we cannot cover here,  but we will make some brief comments on the
  relation with the velocity cut formalism in
  section~\ref{sec:velocitycuts}.\\[10pt]
Key to progress in this interdisciplinary field of scattering amplitudes and classical gravity is that long-distance particle exchanges of gravitons between matter lines can be uniquely tied to observables in general relativity. This allows for on-shell amplitude methods to yield new ways of efficient computation in classical general relativity, thus prompting the radical viewpoint of classical general relativity being suitably defined perturbatively from a path-integral loop expansion. What at first may have looked like a technicality and perhaps a promising new research direction has turned out to lead to an advancement of our understanding of quantum field theory itself. Particularly surprising is the manner in which competing 
factors of Planck's constant $\hbar$ conspire to combine into purely
classical observables. The discovery of this fundamental property of
the quantum field theoretic loop expansion has been prompted by early
derivations of the classical two-body gravitational interactions from
amplitudes~\cite{Iwasaki:1971vb} and only 
recently systematized to all loop
order~\cite{Holstein:2004dn,Bjerrum-Bohr:2018xdl,Kosower:2018adc}.
Interestingly, although we shall seek only classical observables,
quantum mechanical unitarity will be seen to play a crucial
role. Combined with integrand localization on what has been dubbed
{\it velocity cuts}~\cite{Bjerrum-Bohr:2021vuf}  and an exponential
representation of amplitudes in the semi-classical limit it provides a
most efficient procedure for the extraction of the classical
long-distance parts of scattering amplitudes.  An alternative
  approach, with certain similarities,  is  based on an
  heavy-mass effective field theory
  expansion~\cite{Brandhuber:2021eyq}.   An alternative track follows from the evaluation of expectation values~\cite{Kosower:2018adc} rather than a computation of a scattering amplitude {\em per se}. Finally, the eikonal formalism, based on the exponentiation of the gravitational scattering amplitude in impact parameter space~\cite{Amati:1993tb,KoemansCollado:2019ggb,Bern:2020gjj,Cristofoli:2020uzm,DiVecchia:2020ymx,DiVecchia:2021ndb,DiVecchia:2021bdo},
provides an independent procedure for separating classical physics from quantum mechanical scattering amplitudes. The existence of these different field theoretic avenues 
highlights the richness of the problem at hand. In fact, it is
remarkable that Einstein's classical theory of general relativity
turns out to be more easily solved (at least for the purpose of the
post-Minkowskian expansion) by all these different quantum field
theoretic methods than by solving the classical equations of motion
directly. Even more surprisingly, several of the
mathematical structures known from the quantum mechanical loop
amplitudes do survive upon taking the classical limit, thus implying a
deep connection between solutions to the differential equations from
Einstein's non-linear field equations and these mathematical
structures.
This would presumably only with great difficulty be understandable without the link to quantum mechanical scattering amplitudes and their associated loop expansions.
\\[10pt]
The amplitude approach to gravitational scattering of two black holes ignores what can be called the internal structure of black holes. We do not seek a quantum mechanical
description of two black holes interacting gravitationally. Because only large-distance scattering (and a weak coupling expansion in Newton's constant $G_N$) will be considered, 
black holes are treated
as point-like massive objects without a horizon. This is a simple
observation and there is no need to invoke more complicated
argumentation for this obvious separation of scales. We shall always
imagine the two black holes as being separated by enormous distances
while scattering off each other. Nevertheless, once this scattering regime is under control
\cite{Kalin:2019rwq,Bjerrum-Bohr:2019kec} one can attempt to approach
also the bound state regime through the effective
potential~\cite{Kalin:2019inp} by analytic continuation.
The scattering angle, which has  been derived in many
independent methods in different regimes~\cite{Bern:2019crd,Parra-Martinez:2020dzs,DiVecchia:2021ndb,DiVecchia:2021bdo,Herrmann:2021tct,Kalin:2019rwq,Kalin:2019inp,Kalin:2020fhe,Dlapa:2021npj}, is the main link for connecting the scattering
amplitudes to the dynamics of the two-body system.
The Effective One-Body (EOB)
formalism~\cite{Buonanno:1998gg,Buonanno:2000ef} has historically
proven remarkably efficient in extending the validity range of the
post-Newtonian expansion. It is reassuring that this EOB formalism is
robust under an extension to the post-Minkowskian
regime~\cite{Damour:2016gwp,Damour:2017zjx}. One will feed in this
formalism the exact expressions from
the scattering amplitudes valid in all regimes of energy.  Being clearly coordinate-dependent there is much freedom in choosing such an effective one-body metric. From our viewpoint, it is interesting that an EOB formalism exists where an energy-dependent one-body metric exactly reproduces the scattering angle up to third post-Minkowskian order while at the same time being determined immediately by the effective potential of the two-loop scattering amplitude~\cite{Damgaard:2021rnk}. As the accuracy of the post-Minkowskian increases, it becomes phenomenologically relevant to explore the consequences of post-Minkowskian EOB formalisms in terms of gravitational wave predictions.
The amplitude formalism
is indeed a natural framework for investigating  the most fundamental principles of Einstein's
theory of gravity while at the same time establishing new predictions that can be tested by observations.
\section{Classical gravitational scattering from quantum field theory}
\subsection{Einstein's theory of gravity coupled to scalars}
Our starting point is the Einstein-Hilbert action of gravity coupled to matter through the energy-momentum tensor $T_{\mu\nu}$,
\begin{equation}
{\cal S} = \int d^4 x \sqrt{-g} \Bigg[\frac{R}{16 \pi G_N} + g^{\mu\nu} T_{\mu\nu}\Bigg]\,.
\end{equation} 
Here, Newton's constant is denoted by $G_N$, the Ricci scalar is $R$, and we define a weak field expansion around the flat Minkowski space-time metric $\eta_{\mu\nu}$ by  
\begin{equation}
g_{\mu\nu}(x)\equiv \eta_{\mu\nu} + \sqrt{32 \pi G_N}h_{\mu\nu}(x)\,.
\end{equation}
For scalar fields $\phi(x)$, we have the minimal stress-energy tensor 
\begin{equation}
T_{\mu\nu} \equiv \partial_\mu \phi\, \partial_\nu \phi 
- \frac{\eta_{\mu\nu}}{2}\left(\partial^\rho\phi\partial_\rho\phi-m^2 \phi^2\right).
\end{equation}
We shall only be concerned with the scattering of two massive scalar fields coupled to gravity labelled by 
\begin{equation}\varphi_1(p_1,m_1),\ \varphi_2(p_2,m_2) \ \to \
  \varphi_1(p_1',m_1), \ \varphi_2(p_2',m_2),
\end{equation}
where incoming momenta have been denoted by $p_i$ and outgoing momenta by
$p_i'$ with $i=1,2$. The on-shell conditions are
${p_i}^2={p_i'}^2=m_i^2$. \\[10pt]
Now comes an interesting observation. We know that general relativity is a non-linear theory where all physical quantities (such as scattering angles, 
periastron shifts, time delays, etc.) do not truncate at linear order in $G_N$. Let us here focus on the scattering angle.
For two-body gravitational scattering in the post-Minkowskian expansion we must thus compute
the two-to-two scattering amplitude in an expansion in $G_N$. To leading order this is given by a one-graviton exchange as shown in the first diagram below\\[10pt]
\begin{equation}
\begin{gathered}
  \begin{fmffile}{other91xx1}
 \begin{fmfgraph*}(71,51)
\fmfleftn{i}{3}
\fmfrightn{o}{3}
\fmfv{decor.shape=circle,decor.size=(0.00w)}{v3}
\fmf{plain,tension=1}{i1,v1}
\fmf{plain,tension=1}{v2,i3}
\fmf{plain,tension=1}{v2,o3}
\fmf{plain,tension=1}{v1,o1}
\fmf{dbl_wiggly,tension=0}{v1,v3,v2}
\fmf{phantom}{i2,v3,o2}
\end{fmfgraph*}
\end{fmffile}
\end{gathered}
\begin{gathered}
  \begin{fmffile}{other91xx2}
 \begin{fmfgraph*}(71,51)
\fmfleftn{i}{3}
\fmfrightn{o}{3}
\fmfv{decor.shape=circle,decor.size=(0.05w)}{v3}
\fmf{plain,tension=1}{i1,v1}
\fmf{plain,tension=1}{v1,v4}
\fmf{plain,tension=1}{v4,o1}
\fmf{plain,tension=1}{v2,i3}
\fmf{plain,tension=1}{v2,o3}
\fmf{phantom,tension=0}{v1,v2}
\fmf{phantom}{i2,v3}\fmf{phantom}{o2,v3}
\fmf{dbl_wiggly,tension=0}{v1,v3}
\fmf{dbl_wiggly,tension=0}{v4,v3}
\fmf{dbl_wiggly,tension=0}{v2,v3}
\end{fmfgraph*}
\end{fmffile}
\end{gathered}
\ \ \ \
\begin{gathered}
  \begin{fmffile}{other91xx3}
 \begin{fmfgraph*}(71,51)
\fmfleftn{i}{3}
\fmfrightn{o}{3}
\fmfv{decor.shape=circle,decor.size=(0.05w)}{v3}
\fmf{plain,tension=1}{i1,v1}
\fmf{plain,tension=1}{v1,v4}
\fmf{plain,tension=1}{v4,v5}
\fmf{plain,tension=1}{v5,o1}
\fmf{plain,tension=1}{v2,i3}
\fmf{plain,tension=1}{v2,o3}
\fmf{phantom,tension=0}{v1,v2}
\fmf{phantom}{i2,v3}\fmf{phantom}{o2,v3}
\fmf{dbl_wiggly,tension=0}{v1,v3}
\fmf{dbl_wiggly,tension=0}{v4,v3}
\fmf{dbl_wiggly,tension=0}{v5,v3}
\fmf{dbl_wiggly,tension=0}{v2,v3}
\end{fmfgraph*}
\end{fmffile}
\end{gathered}
\ \ \ \ \
\begin{gathered}
  \begin{fmffile}{other91xx4}
 \begin{fmfgraph*}(71,51)
\fmfleftn{i}{3}
\fmfrightn{o}{3}
\fmfv{decor.shape=circle,decor.size=(0.05w)}{v3}
\fmf{plain,tension=1}{i1,v1}
\fmf{plain,tension=1}{v1,v4}
\fmf{plain,tension=1}{v4,v5}
\fmf{plain,tension=1}{v5,o1}
\fmf{plain,tension=1}{v2,i3}
\fmf{plain,tension=1}{vv2,v2}
\fmf{plain,tension=1}{vv2,vvv2}
\fmf{plain,tension=1}{vvv2,o3}
\fmf{phantom,tension=0}{v1,v2}
\fmf{phantom}{i2,v3}\fmf{phantom}{o2,v3}
\fmf{dbl_wiggly,tension=0}{v4,v3}
\fmf{dbl_wiggly,tension=0}{v5,v3}
\fmf{dbl_wiggly,tension=0}{vv2,v3}
\fmf{dbl_wiggly,tension=0}{v2,v1}
\end{fmfgraph*}
\end{fmffile}
\end{gathered}
\end{equation}\\[10pt]
In the non-relativistic limit this of course just gives rise to the Newtonian potential. Although the theory appears linear at this level, the scattering
angle is a non-trivial function of $G_N$ (it is the famous $\arctan$-formula of Newtonian scattering) which will have an infinite-order expansion
in $G_N$. But we do know from elementary considerations of Einstein's equations of motion that there must be corrections to that formula. How can such
corrections appear if we compute the scattering in terms of Feynman diagrams? There is only one possibility:
Without gravitational radiation it is immediately evident that this must entail the computation of
scalar four-point $L$-loop scattering processes (with $L$ arbitrarily large). There are simply no other ways to increase the power of $G_N$ 
while keeping fixed the number of external legs at four for the
two-body problem. So there must be classical contributions to the scattering process residing in the loops!
This simple argument explains why loop amplitudes for gravity
necessarily must contain classical pieces despite the folk-theorem that claims loops to be of
quantum origin only  (see~\cite{Iliopoulos:1974ur}). We shall return to this important point in some detail below where we shall identify precisely those parts of the loop integrations
that give rise to eventually classical contributions to the scattering.\newpage\noindent
For the scattering matrix ${\mathcal M}(p_1,p_2,p_1',p_2')$ we use the following conventions. We denote incoming momenta  by $p_1$ and $p_2$ and outgoing moment by
$p_1'$ and $p_2'$.
The  $\gamma$-factor of the relative velocity is related to the momenta through
by\footnote{For reasons that are obscure to us it
  has become common in the gravity-amplitude community to denote
  the $\gamma$-factor of the relative velocity by $\sigma$. Here we use the normal notation $\gamma$.}
\begin{equation}
\gamma  ~\equiv~ {{p_1\cdot p_2}\over{m_1m_2}}\,.
\end{equation} 
The invariant momentum transfer is defined as usual by
\begin{equation}
q^2 ~\equiv~ (p_1-p_1')^2\equiv(p_2-p_2')^2\,,
\end{equation}
and we also introduce the center-of-mass energy
\begin{equation}\label{e:ECM}
{\cal E}_{CM}^2\equiv(p_1+p_2)^2
\equiv({p_1'}+{p_2'})^2=m_1^2+m_2^2+2m_1m_2\gamma\,.
\end{equation}
With these conventions we have
\begin{equation}
q\cdot
p_1=-q\cdot p_1'={q^2\over2}\,.
\end{equation}
We schematically define the four-point amplitudes ${\mathcal M}(\gamma,q^2)$
expanded into loop amplitudes ${\mathcal M}_L(\gamma,q^2)$ of order $G_N^{L+1}$ by
\begin{equation}
\begin{gathered}
  \begin{fmffile}{other90}
 \begin{fmfgraph*}(101,101)
\fmfleftn{i}{2}
\fmfrightn{o}{2}
\fmfv{decor.shape=oval, decor.filled=shaded, decor.size=(.51w)}{v1}
\fmfv{decor.shape=circle,decor.filled=gray25,decor.size=(.51w)}{v1}
\fmf{fermion,label=$p_1$,label.side=left}{i1,v1}
\fmf{fermion,label=$p_1'$,label.side=left}{v1,i2}
\fmf{fermion,label=$p_2'$,label.side=right}{v1,o2}
\fmf{fermion,label=$p_2$,label.side=right}{o1,v1}
\end{fmfgraph*}
\end{fmffile}
\end{gathered}
=
{\mathcal M}(\gamma,q^2)=
\sum_{L=0}^{ \infty} \mathcal{M}_{ L}(\gamma, q^2) 
\,.
\end{equation}
\subsection{Graviton unitarity cuts}
In principle, loop amplitudes for scalars interacting gravitationally can be straightforwardly enumerated in terms of a standard Feynman diagram expansion. 
In practice,
such a brute-force approach was abandoned long ago, and instead we
today rely almost exclusively on generalized unitarity methods
(see~\cite{Bern:2002kj,Bern:2011qt} for some reviews). Although quantum mechanical
in nature, unitarity more generally can be used to relate products of tree amplitudes to amplitudes of loops. This can be done in stages, from an ordinary Cutkosky
cut to generalized unitarity cuts where more and more internal propagators are taken on-shell.
Such generalized unitarity methods are able to produce, step-by-step, all those contributions 
from loop amplitude which are ``cut
constructible''~\cite{Bern:1994cg,Bern:2007xj}. All non-analytic elements of loop amplitudes can be obtained in this manner (and the remaining rational
terms can be extracted with a bit more effort).
Generalised unitarity thus provides an immediate simplification of amplitude computations. In fact, most precision calculation of scattering cross sections in 
the Standard Model of particle physics today rely on this. Most importantly for the present purposes is the fact that in order to obtain the long-distance 
classical gravity contributions to the scattering of two
heavy objects we should discard all analytical pieces. Those terms correspond to ultra-local pieces after a Fourier transform. This means that generalized
unitarity methods are almost ideally suited for the calculation of classical gravitational scattering. Physically, we can understand this from the fact that
long-distance gravitational effects literally can be viewed as coming from those parts of the loop diagrams where the exchanged (virtual) gravitons are almost on mass
shell and thus propagating over large distances. For massive scattering, these parts of the amplitude are functions of dimensionless
ratios such as $m/\sqrt{-q^2}$, where $m$ is a heavy mass scale.\\[10pt]
Let us illustrate our way of using unitarity to reconstruct the
non-analytical parts of the four-point scattering amplitude. An $(L+1)$-graviton cut 
is defined by\footnote{We are primarily interested in
Einstein's gravity in $D=4$ but we use dimensional regularization to tame divergences at intermediate stages.}
\begin{multline}\label{e:MLcut}
 i\mathcal {{}\cal M}_{{} L+1}^{{}\textrm{{}cut}} (\gamma,q^2)\equiv\hbar^{{}3L+1} \int (2\pi)^D\delta(q+\ell_2+\cdots+\ell_{{}L+2})\prod_{{}i=2}^{{}L+2} {{}i
 \over\ell_i^2}\prod_{{}i=2}^{{}L+2} {{} d^D\ell_i\over (2\hbar\pi)^D} 
 \cr{{}1\over (L+1)!}\sum_{{}h_i=\pm2}
 {{}\cal M}^{{}\rm tree}_{{}\rm Left}(p_1,\ell_2^{{}h_2},\ldots,\ell_{{}L+2}^{{}h_{{}L+2}},-p_1')
{{}\cal  M}^{{}\rm tree}_{{}\rm Right}(p_2,-\ell_2^{{}h_2},\ldots,-\ell_{{}L+2}^{{}h_{{}L+2}},-p_2')^\dagger,
\end{multline}
which can be represented by\\[5pt]
\begin{equation}
\hspace{-1.2cm} {\mathcal M}^{\rm cut}_{ L+1}(\gamma,q^2)=
\hspace{.4cm} \begin{gathered}
\begin{fmffile}{Lcut}
\begin{fmfgraph*}(105,105)
\fmfstraight
\fmfleftn{i}{2}
\fmfrightn{o}{2}
\fmftop{t}
\fmfbottom{b}
\fmfrpolyn{smooth,filled=30,label=\textrm{\ tree\ }}{el}{8}
\fmfrpolyn{smooth,filled=30,label=\textrm{tree}}{er}{8}
\fmf{fermion,label=$p_1$,label.side=left,tension=1.8}{i1,el1}
\fmf{fermion,label=$p_1'$,label.side=left,tension=1.8}{el3,i2}
\fmf{fermion,label=$p_2'$,tension=1.8}{er5,o2}
\fmf{fermion,label=$p_2$,label.side=right,tension=1.8}{o1,er7}
\fmf{dbl_wiggly,tension=.12}{el5,er3}
\fmf{dots,tension=0}{el6,er2}
\fmf{dots,tension=0}{el2,er6}
\fmf{dbl_wiggly,tension=.12}{el7,er1}
\fmf{dashes,for=red}{b,t}
\end{fmfgraph*}
\end{fmffile}
 \end{gathered}
\end{equation}\\[5pt]
where the middle dots represent $L-1$ additional cut graviton lines.
The amplitudes  ${\cal M}^{{}\rm tree}_{\rm Left}(p_1,\ell_2,\ldots,\ell_{L+2},-p_1')$ and
${\cal M}^{\rm tree}_{\rm Right}(p_2,-\ell_2,\ldots,-\ell_{L+2},-p_2')$ denote tree-level multi-graviton emission from a massive scalar line. 
The two pieces are glued together with the inserted propagator factors so that indeed the full expression provides the cut of the shown part of
the amplitude stemming from only graviton exchanges between the two massive legs.
Note that we have universal conventions with all graviton lines incoming in the left tree factor and out-going in the right tree factor, $i.e.$ 
\begin{equation}
 q=p_1-p_1'=-p_2+p_2'=-\sum_{i=2}^{L+2}\ell_i\,. 
\end{equation}
These shown cut diagrams clearly describe only a subset of all diagrams for the full amplitude. We can understand intuitively from the above arguments that
these should contain the bulk of the long-distance non-analytical parts. Indeed, it was shown in refs.~\cite{Bjerrum-Bohr:2021vuf,Bjerrum-Bohr:2021din} 
that the classical contributions from the full amplitude at one and two loop level can be calculated from these $(L=1)$ and $(L=2)$ cuts {\em except} for 
a small set of diagrams that are not amenable to three-graviton cuts. These diagrams, which happen to vanish in maximal supergravity, must be included
if we want the complete classical part of the amplitude in Einstein gravity. They correspond to self-energy and vertex corrections (some, leaping over
two vertices are denoted as ``mushroom diagrams'') and share the common property of always involving a graviton exchange from one massive line back to itself. 
Physically, this will clearly correspond to what can be considered radiation-reaction terms: it is the gravitational field reacting back on the same
scalar line. However, not all radiation-reaction terms come from this class of diagrams, a class which, as mentioned above, vanishes in maximal supergravity. 
The other pieces
come from the three-graviton cut, once all classical contributions from the integrations are properly included. This was first realized by a beautiful
argument based on analyticity and crossing symmetry in ref.~\cite{DiVecchia:2020ymx} and later verified by the first explicit computation of the full
classical part of the two-loop amplitude in ref.~\cite{Bjerrum-Bohr:2021vuf}. We shall discuss this in much greater detail below.
\section{The classical potential from a Lippmann-Schwinger equation }
It is a classical problem in scattering theory to relate
the scattering amplitude $\mathcal M$ to an interaction potential
$\mathcal V$. This is typically phrased in terms of non-relativistic quantum mechanics, but it is readily generalized to the relativistic case. Crucial in this respect is the fact that we shall consider particle solutions to the relativistic equations only.
There will thus be, in the language of old-fashioned (time-ordered) perturbation theory, no back-tracking diagrams corresponding to multi-particle intermediate states. This is trivially so since we neither wish to treat the macroscopic classical objects such as heavy neutron stars as indistinguishable particles with their corresponding antiparticles nor do we wish to probe the scattering process in any potential annihilation channel. The classical objects that scatter will always be restricted to classical distance scales. The proper language for the Hamiltonian $H$ and hence potential $\mathcal{V}$ is therefore the so-called Salpeter equation for the two massive
scalars. In the center-of-mass frame with three-momenta $p = |\vec{p}_1| = |\vec{p}_2|$,
\begin{equation}\label{e:H}
H ~=~ \sqrt{p^2 + m_1^2} + \sqrt{p^2 + m_2^2} + {\mathcal V}(r,p)\,,
\end{equation}
which indeed, as seen, excludes antiparticles of the massive states.
\\[10pt]
It is convenient to introduce the amplitude $\tilde{\mathcal{M}}$ in the non-relativistic normalization convention
\begin{equation}
\tilde{\mathcal{M}}(p,p') ~\equiv~ \frac{\mathcal{M}(p,p')}{4E_1E_2}\,,
\end{equation}
with $p_1=(E_1,\vec p)$, $p_1'=(E_1, \vec p\,')$, $p_2=(E_2, -\vec p)$ and $p_2'=(E_2, -\vec p\,')$.
As a first observation one notices that to leading order, and in the non-relativistic limit, the classical potential is
simply equal to the amplitude after a Fourier transform:
\begin{equation}
    \mathcal V(r,p)= \int {d^3q\over(2\pi)^3} e^{i q\cdot r} \mathcal V(p,q) = \int {d^3q\over(2\pi)^3} e^{i q\cdot r} \tilde{\mathcal{M}}(p,q)\,. ~~~~~~~{\rm (Tree~ level)}
  \end{equation}
The extension of this relationship beyond tree-level was established long ago in terms of the Born series. It is most succinctly phrased in terms
the Lippman-Schwinger equation~\cite{Cristofoli:2019neg} that relates the scattering amplitude $\mathcal M$ and the potential $\mathcal V$ in an exact closed form: 
\begin{equation}
  \tilde{\mathcal M}(p,p') = \mathcal V(  p,p')+ \int
  {d^3k\over (2\pi)^3}   { \mathcal V(
    p,k) \mathcal M(k,p')\over E_p-E_k+i\varepsilon}\,, \label{Lippmann}
\end{equation}
which is given here in momentum space. This framework is a relativistic extension, through a one-particle
Hamiltonian and the associated Salpeter equation, of the conventional approach to determining the interaction
potential in perturbation gravity by means of Born subtractions. The
Lippmann-Schwinger equation summarizes this procedure in a very clear form and it furnishes in a transparent and systematic manner the needed
Born subtractions at arbitrary loop order. 

The Lippmann-Schwinger approach is equivalent to the Effective Field Theory approach introduced in ref.~\cite{Cheung:2018wkq} and pursued in the two-loop
calculation of refs.~\cite{Bern:2019nnu,Bern:2019crd}. This is quite easily shown~\cite{Cristofoli:2020uzm}. First, we can solve the Lippmann-Schwinger equation~\eqref{Lippmann} perturbatively in the potential $\mathcal{V}$ to get
\begin{multline}
\mathcal{{\tilde{M}}}(\vec p,\vec p\,')=\mathcal{V}(\vec p,\vec p\,')\cr
+\sum_{n=1}^{\infty}\int \frac{d^{D-1} \vec k_1}{(2\pi \hbar)^{D-1}} \, \frac{d^{D-1}\vec k_2}{(2\pi \hbar)^{D-1}}  \cdots  \frac{d^{D-1} \vec k_n}{(2\pi \hbar)^{D-1}} 
\frac{\mathcal{V}(\vec p,\vec k_1) \cdots \mathcal{V}(\vec k_{n},\vec p\,')}{(E_{p}-E_{k_1}+i \epsilon) \cdots(E_{k_{n-1}}-E_{k_{n}}+i \epsilon)}\,,
\end{multline}
or, by expanding both $\mathcal{{\tilde{M}}}$ and $\mathcal{V}$ in powers of $G_N$ and, for illustration, truncating at one-loop order,
\begin{equation}
\mathcal{V}_{\rm 1PM}(\vec p,\vec p\,')+
\mathcal{V}_{\rm 2PM}(\vec p,\vec
p\,')=\mathcal{\tilde{M}}_{\rm tree}(\vec p,\vec
p\,')+\mathcal{\tilde{M}}_{\rm 1-loop}(\vec p,\vec p\,')+
\mathcal{\tilde{M}}_{\rm B}(\vec p,\vec p\,')\,, \label{Vexp}
\end{equation}
where the first Born subtraction is given by
\begin{equation}
\mathcal{\tilde{M}}_{\rm B}(\vec p,\vec p\,') ~\equiv~
-\int\frac{d^{D-1}\vec
  k}{(2\pi\hbar)^{D-1}}\frac{\mathcal{\tilde{M}}_{\rm tree}(\vec p,\vec k)
\mathcal{\tilde{M}}_{\rm tree}(\vec k,\vec p\,')}{E_p-E_k + i\varepsilon}\,.
\end{equation}
The solution at one-loop level is thus found recursively as illustrated above and eq.~\eqref{Vexp} provides the one-loop potential ${\mathcal V}_{\rm 2PM}$.

The effective field theory approach for classical gravity~\cite{Cheung:2018wkq} is based on a different principle but the resulting equation for the potential
is the same. The starting point of this effective field theory
formalism is the parametrization of the potential $\mathcal{V}$ in an operator basis, 
\begin{equation}
\mathcal{V}(\vec p, \vec p\,')= 
G_N \,c_1\left(\frac{p^2+p'^2}{2}\right)\times \left(\frac{q^2}{\hbar^2}\right)^{D-5}+ G_N^2\, c_2\left(\frac{p^2+p'^2}{2}\right)\times \left(\frac{q^2}{\hbar^2}\right)^{\frac{D-5}{2}}+\cdots\,,
\end{equation}
including terms here up to one-loop order. Here $c_i$-coefficients (chosen here to depend symmetrically on the momenta, as shown) are to be fixed by a matching condition between the Effective Field Theory and the
fundamental underlying gravity theory. At $L$-loop order it reads
\begin{equation}
\mathcal{\tilde{M}}_{\rm L-loop}(\vec p,\vec p\,')=\mathcal{M}_{\rm (L+1)PM}^{EFT}(\vec p,\vec p\,') \,,   
\end{equation}
which at tree level reduces to
\begin{equation}
\mathcal{\tilde{M}}_{\rm tree}(\vec p, \vec
p\,')=\mathcal{M}^{\rm EFT}_{\rm 1PM}(\vec p, \vec p\,')\,.
\end{equation}
Since in the effective theory the first interaction term is the contact term given in $D=4$ dimensions by 
\begin{equation}
\mathcal{V}(\vec p, \vec p\,')= 
G_N \,c_1\left(\frac{p^2+p'^2}{2}\right)\times \left(\frac{q^2}{\hbar^2}\right)^{-1}\,,
\end{equation}
this fixes $c_1$ to be what we already knew: the tree-level potential
is given entirely by the tree-level amplitude whose precise form is recalled below. So at this stage the first coefficient $c_1$ has been fixed.
At one-loop order, the matching condition
remains unchanged but when expanded in $G_N$ there will be two contributions to the Effective Field Theory amplitude: one from the new vertex determined by
the so far unknown $c_2$-coefficient, another from the loop (bubble) contribution from the leading-order interaction. The matching condition
\begin{equation}
 {\mathcal{\tilde{M}}}_{\rm 1-loop}(\vec p, \vec
 p\,')={\mathcal{M}}^{\rm EFT}_{\rm 2PM} (\vec p, \vec p\,')\,,
\end{equation}
is easily shown to give
\begin{equation}
\label{levelmatch}
\mathcal{\tilde{M}}_{\rm 1-loop}(\vec p,\vec p\,') = \mathcal{V}_{\rm 2PM}(\vec p,\vec p\,')
- \widetilde{\mathcal M}_{\rm B}(\vec p, \vec p\,')\,.
\end{equation}
We learn that the first Born subtraction term $\widetilde{\mathcal M}_{\rm B}(\vec p, \vec p\,')$ is simply the bubble graph of the one-loop Effective Field Theory. Moreover, the
new coefficient $c_2$ has now been fixed by the remaining parts of the one-loop amplitude. This equivalence between the Lippmann-Schwinger and Effective Field Theory
approaches is easily shown to hold to all orders~\cite{Cristofoli:2020uzm}.\\[10pt]
Neither the Lippmann-Schwinger approach nor the Effective Field Theory approach as outlined here have been pursued beyond two-loop order. Instead,
alternative strategies have been pursued that are closer in spirit to the eikonal formalism. Nevertheless, the subtraction schemes that have been
learned from the Lippmann-Schwinger and Effective Field Theory methods do have analogues in the alternative methods -- in interesting ways. This will be discussed below.
\section{The $\hbar$ expansion from the exchange of gravitons between matter lines}\label{sec:perPM}
At a given order in perturbation theory there are exchanges of gravitons
(curly lines) between massive external matters (solid lines)
$$\includegraphics[width=14cm]{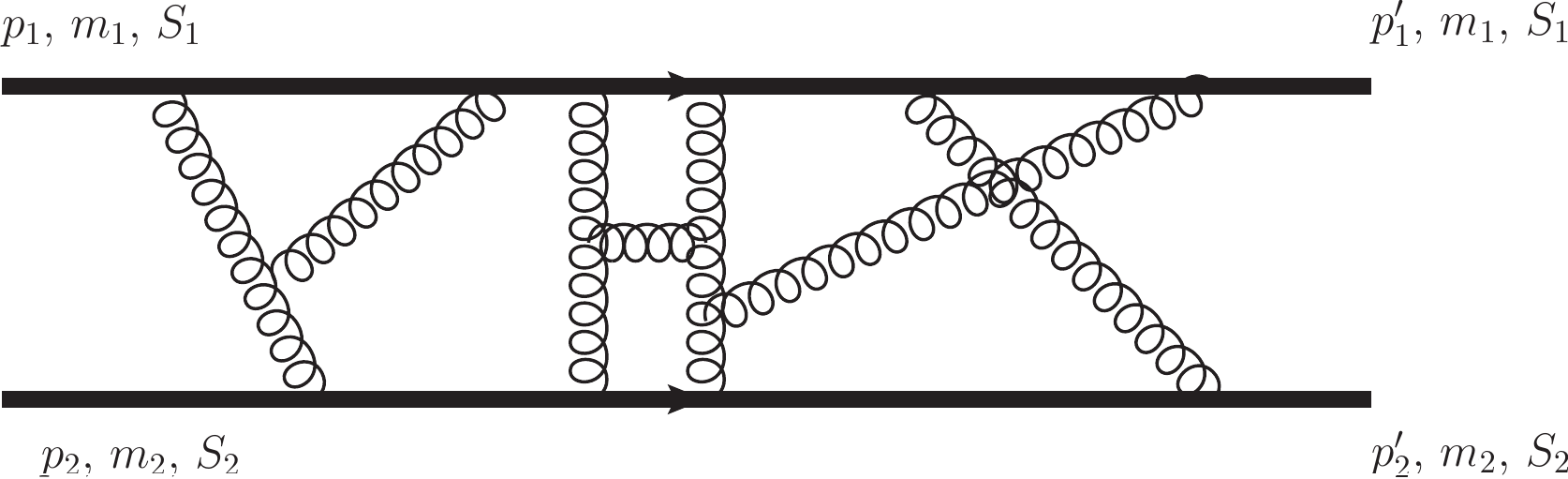}$$
as well as diagrams with gravitons beginning and terminating on the same matter lines. Graviton self-energy diagrams can be discarded since they will never contribute to the 
classical result: even if included, their contributions will always be cancelled when computing the scattering angle.\\[10pt]
A standard textbook argument states that the $L$-loop  contribution should be of order 
$ \mathcal M_L(\gamma,q^2)=\mathcal O(\hbar^{L-1} )$~\cite{Iliopoulos:1974ur}. 
However, a different scaling emerges when keeping
the wave-number $\underline q= q/\hbar$ fixed and taking both the
$\hbar\to0$ and the small momentum transfer $q\to0$
limits~\cite{Holstein:2004dn,Bjerrum-Bohr:2018xdl,Kosower:2018adc}.
The $L$-loop two-body scattering amplitude has a Laurent expansion
around four dimensions~\cite{Bjerrum-Bohr:2021vuf}
\begin{equation}\label{e:Mhbarexp}
     \mathcal M_L(\gamma,\underline q^2,\hbar)={\mathcal
       M_L^{(-L-1)}(\gamma,\underline q^2)\over
       \hbar^{L+1}|\underline q|^{{L(4-D)\over2}+2}}+\cdots +{\mathcal
       M^{(-1)}_L(\gamma,\underline q^2)\over \hbar |\underline q|^{{L(4-D)\over2}+2-L}}+O(\hbar^0)\,.
   \end{equation}
The part of the amplitude that will contribute to the
classical interactions depicted in the above figure is the contribution of order $1/\hbar$.
The full quantum amplitude contains three types of contributions: (1) a
term of order $1/\hbar^r$ with $2\leq r\leq L+ 2$ that are more
singular than the classical piece, (2) a classical piece of order
$1/\hbar$ and (3) quantum corrections of order $\hbar^r$ with
$r\geq0$. All these contributions are constrained (and in fact dictated) by unitarity of
the $S$-matrix, as we will explain in section~\ref{sec:exp}.
\subsection{Tree level and one-loop amplitudes in Einstein gravity}
The tree-level amplitude is given by\\[5pt]
\smallskip
\begin{equation}\label{e:tree}
  {\mathcal M}_0(\gamma,\underline q^2,\hbar)=
  \begin{gathered}\begin{fmffile}{treegraph}\begin{fmfgraph*}(80,40)
      \fmfleftn{i}{2}
      \fmfrightn{o}{2}
\fmf{plain}{o1,v3,o2}
\fmf{plain}{i1,v1,i2}
\fmflabel{$p_1$}{i1}
\fmflabel{$p_1'$}{i2}
\fmflabel{$p_2'$}{o2}
\fmflabel{$p_2$}{o1}
\fmf{dbl_wiggly,tension=.6,label=$q^2$}{v1,v3}
\end{fmfgraph*}
\end{fmffile}
\end{gathered}=\hbar {2\pi
       m_1^2m_2^2  G_N(2\gamma^2-1)\over
       |\underline q|^2}+O(\hbar^0)\,.
\end{equation}\\[5pt]
In refs.~\cite{Bjerrum-Bohr:2013bxa, Bjerrum-Bohr:2021vuf} the one-loop two-body scattering amplitude was computed by means of two-particle cuts (in the following we employ the notation in those papers)
\begin{equation}
 {\mathcal M}_1(\gamma,\underline q^2,\hbar)=
\qquad  \begin{gathered}
 \begin{fmffile}{2cut}
 \begin{fmfgraph*}(101,101)
\fmfstraight
\fmfleftn{i}{2}
\fmfrightn{o}{2}
\fmftop{t}
\fmfbottom{b}
\fmfrpolyn{smooth,filled=31,label=\textrm{tree}}{el}{8}
\fmfrpolyn{smooth,filled=31,label=\textrm{tree}}{er}{8}
\fmf{fermion,label=$p_1{}$,label.side=left,tension=2}{i1,el1}
\fmf{fermion,label=$p_1'{}$,label.side=left,tension=2}{el3,i2}
\fmf{fermion,label=$p_2'{}$,tension=2}{er5,o2}
\fmf{fermion,label=$p_2{}$,label.side=right,tension=2}{o1,er7}
\fmf{dbl_wiggly,tension=.101}{el5,er3}
\fmf{dbl_wiggly,tension=.101}{el7,er1}
\fmf{dashes,for=red}{b,t}
\end{fmfgraph*}
\end{fmffile}
\end{gathered} \qquad .
\end{equation}
The one-loop amplitude can be decomposed terms of master integrals as follows
\begin{equation}\label{e:M1total}
 \mathcal M_1(\gamma,\underline q^2,\hbar )=\mathcal M_1^{\Box}{}+\mathcal
 M_1^{\triangleright}{}+\mathcal M_1^{\triangleleft}{}+\mathcal M_1^{\circ}\,,
\end{equation}
with coefficients provided by the two-graviton unitarity cut. The amplitude has a Laurent expansion in $\hbar$ as explained above. Including the first quantum correction it reads
\begin{equation}\label{e:M1expand}
  \mathcal M_1(\gamma,\underline q^2,\hbar)={1\over
        |\underline q|^{4-D}}\left({ \mathcal M_1^{(-2)}(\gamma,\underline q^2)\over\hbar^2}+ { \mathcal
        M_1^{(-1)}(\gamma,\underline q^2)\over\hbar}+  \mathcal M_1^{(0)}(\gamma,\underline q^2)+\mathcal O(\hbar)\right)\,,
\end{equation}
with
\begin{align}
  \mathcal M_1^{(-2)}(\gamma,\underline q^2)&=\mathcal M_1^{\Box(-2)}(\gamma,\underline q^2),\cr
                       \mathcal M_1^{(-1)}(\gamma,\underline q^2)&=\mathcal
                                            M_1^{\Box(-1)}(\gamma,\underline q^2)+\mathcal
                                            M_1^{\triangleright(-1)}(\gamma,\underline q^2)+\mathcal
                                            M_1^{\triangleleft(-1)}(\gamma,\underline q^2),\cr
 \mathcal M_1^{(0)} (\gamma,\underline q^2)&=\mathcal M_1^{\Box(0)}(\gamma,\underline q^2)+\mathcal
                                            M_1^{\triangleright(0)}(\gamma,\underline q^2)+\mathcal
                                            M_1^{\triangleleft(0)}(\gamma,\underline q^2)+\mathcal
                     M_1^{\circ(0)}(\gamma,\underline q^2)\,.
\end{align}
Defining amplitudes in $b$-space by a suitably normalized Fourier transform,
\begin{equation}
  \widetilde{\mathcal M}(\gamma,b,\hbar)=\frac{1}{4 m_1m_2\sqrt{\gamma^2-1}}\int_{\mathbb R^{D-2}} \frac{d^{D-2}\vec{\underline
      q}}{(2\pi)^{D-2}}\mathcal M(\gamma,\underline q^2,\hbar) e^{i
    \vec{\underline q}\cdot\vec{b}}\,,
\end{equation}
we find the classical and leading quantum pieces
from $ \mathcal M_1^{(-1)}$ and $  \mathcal M_1^{(0)}$, respectively. The classical part is
\begin{equation}\label{e:Mone}
\widetilde{\mathcal M}_{1{}}^{\rm Cl.}(\gamma,b,\hbar)=\frac{3 \pi G_N^2 (m_1+m_2) m_1 m_2 (5\gamma^2-1)}{4 b \sqrt{{}\gamma^2-1}\hbar} (\pi b^2 e^{\gamma_E})^{4-D}+{}\mathcal O(4-D)\,,
\end{equation}
while the leading quantum correction reads 
\begin{multline}
 \widetilde{\mathcal M}_{1}^{{}\rm Qt.}(\gamma,b)={}\frac{G_N^2 (\pi b^2
   e^{\gamma_E})^{4-D} }{b^2}\Bigg(i {4-D\over2}\frac{
  (2\gamma^2-1)^2\mathcal E_{\rm C.M.}^2}{(\gamma^2-1)^2{}}\cr
 -\frac{m_1 m_2 }{\pi
   (\gamma^2-1{})^{\frac{3}{2}}}\Big(\frac{1-49\gamma^2+18
   \gamma^4}{15}{}-\frac{{}2\gamma(2\gamma^2-1)(6\gamma^2-7)\arccosh({}{}\gamma)}{\sqrt{{}{}\gamma^2-1}}\Big)
 \Bigg)+\mathcal O( (4-D)^2)\,.
\end{multline}
The dimension-dependent pieces of these amplitudes 
will feed into certain contributions at the next loop order. They are
needed for a proper identification of the classical part of the
two-loop amplitude in~\eqref{e:ResultM2}.

\subsection{The two-loop amplitude for Einstein gravity}
The two-loop amplitude
\begin{equation}\label{e:M2}
  \mathcal M_2(\gamma,q^2)=   \mathcal M_2^{\rm
  3-cut}(\gamma,q^2)+\mathcal M_2^{\rm SE}(\gamma,q^2)\,,
\end{equation}
receives a contribution from the three-particle cut,  $\mathcal M_2^{\rm
  3-cut}(\gamma,q^2)$, with only gravitons propagating across the cut, plus a set of diagrams with a graviton beginning and
terminating on the same matter lines, $\mathcal M_2^{\rm SE}(\gamma,q^2)$. This includes massive self-energy diagrams, vertex corrections, and so-called mushroom diagrams. Their contributions
to the 2-loop amplitude have collectively been denoted by $\mathcal M_2^{\rm SE}(\gamma,q^2)$.
The three particle-cut is evaluated in dimension
$D$, with $D>4$ for regulating  the infrared divergences of the classical part,
\begin{align}\label{e:3cut}
&  \mathcal M_{2}^{\rm 3-cut}(\gamma,\underline q^2,\hbar)=
 \hspace{.4cm}   \begin{gathered}
\begin{fmffile}{3cut}
 \begin{fmfgraph*}(100,100)
\fmfstraight
\fmfleftn{i}{2}
\fmfrightn{o}{2}
\fmftop{t}
\fmfbottom{b}
\fmfrpolyn{smooth,filled=30,label=\textrm{tree}}{el}{8}
\fmfrpolyn{smooth,filled=30,label=\textrm{tree}}{er}{8}
\fmf{fermion,label=$p_1$,label.side=left,tension=2}{i1,el1}
\fmf{fermion,label=$p_1'$,label.side=left,tension=2}{el3,i2}
\fmf{fermion,label=$p_2'$,tension=2}{er5,o2}
\fmf{fermion,label=$p_2$,label.side=right,tension=2}{o1,er7}
\fmf{dbl_wiggly,tension=.1}{el5,er3}
\fmf{dbl_wiggly,tension=.1}{el6,er2}
\fmf{dbl_wiggly,tension=.1}{el7,er1}
\fmf{dashes,for=red}{b,t}
\end{fmfgraph*}
\end{fmffile}
\end{gathered}\\
&=\int \frac{d^D l_1
  d^D
  l_2d^Dl_3}{(2\pi{})^{3D}} (2\pi)^D\delta^{(D)}(l_1+l_2+l_3+q){i^3\over
                   l_1^2l_2^2l_3^2}\cr
    \nonumber               &\times{1\over3!}\sum_{\textrm{Perm}(l_1,l_2,l_3)\atop
  \lambda_1=\pm,\lambda_2=\pm,\lambda_3=\pm} \mathcal M_0(p_1,p_1',l_1^{\lambda_1},l_2^{\lambda_2},l_3^{\lambda_3})(\mathcal M_0(p_2,p_2',-l_1^{\lambda_1},-l_2^{\lambda_2},-l_3^{\lambda_3}))^{*}\,,
\end{align}
which involves two five-point tree-level amplitudes.
The sum is over the physical states across the cut and can conveniently be done employing spinor-helicity variables 
$\lambda_i$ or using expressions for covariant tree amplitudes as outlined in \cite{Bjerrum-Bohr:2021wwt} which is based on the representation of tree amplitudes discussed in refs. \cite{Bjerrum-Bohr:2019nws,Bjerrum-Bohr:2020syg}.  As explained in the introduction, we need only the cut-constructible part of
the amplitude in order to extract the long-range classical contributions.
The $\hbar$-counting established in eq.~(3.6)
of~\cite{Bjerrum-Bohr:2021vuf}, makes it is clear that at least two massive propagators are needed in order to obtain a classical contribution from the amplitude. 
Employing a partial-fraction decomposition of the tree-level amplitudes
with references to the linear propagators $p_1\cdot l_i$ and $p_2\cdot
l_i$ with $i=1,2,3$, the three-particle cut can be reorganized into five specific topologies that contribute to the classical result
\begin{equation}
\mathcal M_{2}^{\rm 3-cut}(\gamma,\underline q^2,\hbar)=\mathcal M_2^{\dBox}+\mathcal
M_2^{\triangleboxleft }+\mathcal M_2^{\triangleboxright}+\mathcal
M_2^{\triangleleft\triangleleft}+\mathcal M_2^{\triangleright\triangleright}+\mathcal
M_2^{H}+\mathcal M_2^{\Box\circ}\,.
\end{equation}
The two-loop amplitude for Einstein gravity  involves integrals with
non-trivial numerators and corresponding to many graph topologies. It is a
remarkable fact that the reduction to master integrals  (for
instance using the automatic integral reduction program {\tt LiteRed}~\cite{Lee:2013mka}) 
involves only nine such basis integrals.\\[10pt]
As mentioned above, the three-particle graviton cut does not provide the full two-loop contribution
to classical gravitational scattering in Einstein gravity.
Contributions from diagrams with one graviton beginning and terminating on the same matter lines are needed as well. We write them as
\begin{equation}
\mathcal M_{2}^{\rm self-energy}(\gamma,\underline q^2)=-4(16\pi G_N)^3\sum_{i=I}^{IV}
(J_{SE}^{i,s}+J_{SE}^{i,u})+(m_1\leftrightarrow m_2)\,,
\end{equation}
where the integrals $J_{SE}^{i,s}$ are given in
eq.~\eqref{e:Jseis}--\eqref{e:Jseivs} below. The contributions
$J_{SE}^{i,u}$ are obtained by the exchange of the legs $p_2$ and
$p_2'$ and there are of course also all the symmetric contributions with a graviton
line beginning and terminating on the massive line of mass $m_2$.
\begin{multline}
  J_{SE}^{I,s}= \begin{gathered}
    \begin{fmffile}{SEI}
    \begin{fmfgraph*}(100,100)
\fmfstraight
\fmfleftn{i}{2}
\fmfrightn{o}{2}
\fmf{fermion,label=$p_1$,label.side=left}{i1,vv1}
\fmf{plain,tension=2}{vv1,v1}
\fmf{fermion,label=$p_1'$,label.side=left}{vv2,i2}
\fmf{plain,tension=2}{v2,vv2}
\fmf{fermion,label=$p_2'$}{vv3,o2}
\fmf{plain,tension=2}{v3,vv3}
\fmf{fermion,label=$p_2$,label.side=right}{o1,vv4}
\fmf{plain,tension=2}{v4,vv4}
\fmf{fermion}{v1,v2}
\fmf{fermion}{v4,v3}
\fmf{dbl_wiggly}{v4,v1}
\fmf{dbl_wiggly}{v2,v3}
\fmf{dbl_wiggly,tension=0}{vv1,vv2}
\fmf{phantom,tension=0}{vv3,vv4}
\end{fmfgraph*}
\end{fmffile}
\end{gathered}  
  = \int \frac{d^D l_1 d^D l_2}{(2\pi)^{2D}}
\frac{ (16\pi G_N)^3 \hbar^7
                }{((p_1-l_1-l_2)^2-m_1^2+i\varepsilon)((p_1-l_2)^2-m_1^2+i\varepsilon)}\cr
                  \times \frac{m_1^8 m_2^4 (2\sigma^2-1)^2+2m_1^6 m_2^4 (2\sigma^2-1)^2 |\hbar\underline q|^2}{((p_1-l_1-l_2-q)^2-m_1^2+i\varepsilon)((p_2-l_1)^2-m_2^2+i\varepsilon)l_1^2 (l_1+q)^2 (l_1+l_2)^2}, \label{e:Jseis} 
\end{multline}
\begin{multline}
J_{SE}^{II,s}=  \begin{gathered} \begin{fmffile}{SEII}
    \begin{fmfgraph*}(100,100)
\fmfstraight
\fmfleftn{i}{2}
\fmfrightn{o}{2}
\fmf{fermion,label=$p_1$,label.side=left}{i1,vvv1}
\fmf{plain,tension=2}{vvv1,v1}
\fmf{fermion,label=$p_1'$,label.side=left}{vv2,i2}
\fmf{plain,tension=2}{v2,vv2}
\fmf{fermion,label=$p_2'$}{vv3,o2}
\fmf{plain,tension=2}{vv3,v3}
\fmf{fermion,label=$p_2$,label.side=right}{o1,vv4}
\fmf{plain,tension=2}{v4,vv4}
\fmf{dbl_wiggly,tension=0.5}{v4,v1}
\fmf{dbl_wiggly,tension=0.5}{v2,v3}
\fmf{plain,tension=2}{mm1,v2}
\fmf{fermion}{m1,mm1}
\fmf{plain,tension=2}{v1,m1}
\fmf{fermion}{mm3,m3}
\fmf{plain,tension=2}{v4,mm3}
\fmf{plain,tension=2}{v3,m3}
\fmf{dbl_wiggly,left,tension=0.}{m1,vv2}
\fmf{phantom,left,tension=0.}{vvv1,vv2}
\fmf{phantom,left,tension=0.}{m3,vv3}
\fmf{phantom,left,tension=0.}{m3,vv4}
\fmf{phantom,left,tension=0.}{m3,mm3}
\fmf{phantom,left,tension=0.}{m1,mm1}
\fmf{phantom,left,tension=0}{vvv1,vv2}
\fmf{phantom,left,tension=0}{vv3,vv4}
\end{fmfgraph*}
\end{fmffile}
\end{gathered} 
               = \int \frac{d^D l_1 d^D l_2}{(2\pi)^{2D}}
\frac{ (16\pi G_N)^3 \hbar^7\left(
m_1^8 m_2^4 (2\sigma^2-1)^2
             \right)    }{((p_1-l_1-l_2)^2-m_1^2+i\varepsilon)((p_1-l_2)^2-m_1^2+i\varepsilon)}\cr
                \times \frac{1}{((p_1+l_1)^2-m_1^2+i\varepsilon)((p_2-l_1)^2-m_2^2+i\varepsilon)l_1^2 (l_1+q)^2 (l_1+l_2)^2}\,,\label{e:Jseiis}
\end{multline}
\begin{multline}\label{e:Jseiiis}
  J_{SE}^{III,s}= \begin{gathered} \begin{fmffile}{SEIII}
    \begin{fmfgraph*}(100,100)
\fmfstraight
\fmfleftn{i}{2}
\fmfrightn{o}{2}
\fmf{fermion,label=$p_1$,label.side=left}{i1,vvv1}
\fmf{plain,tension=2}{vvv1,v1}
\fmf{fermion,label=$p_1'$,label.side=left}{vv2,i2}
\fmf{plain,tension=2}{v2,vv2}
\fmf{fermion,label=$p_2'$}{vv3,o2}
\fmf{plain,tension=2}{vv3,v3}
\fmf{fermion,label=$p_2$,label.side=right}{o1,vv4}
\fmf{plain,tension=2}{v4,vv4}
\fmf{dbl_wiggly,tension=0.5}{v4,v1}
\fmf{dbl_wiggly,tension=0.5}{v2,v3}
\fmf{plain,tension=2}{mm1,v2}
\fmf{fermion}{m1,mm1}
\fmf{plain,tension=2}{v1,m1}
\fmf{fermion}{mm3,m3}
\fmf{plain,tension=2}{v4,mm3}
\fmf{plain,tension=2}{v3,m3}
\fmf{dbl_wiggly,left,tension=0.}{vvv1,mm1}
\fmf{phantom,left,tension=0.}{vvv1,vv2}
\fmf{phantom,left,tension=0.}{m3,vv3}
\fmf{phantom,left,tension=0.}{m3,vv4}
\fmf{phantom,left,tension=0.}{m3,mm3}
\fmf{phantom,left,tension=0.}{m1,mm1}
\fmf{phantom,left,tension=0}{vvv1,vv2}
\fmf{phantom,left,tension=0}{vv3,vv4}
\end{fmfgraph*}
\end{fmffile}
\end{gathered}
                  = \int \frac{d^D l_1 d^D l_2}{(2\pi)^{2D}} \frac{ (16\pi G_N)^3 \hbar^7\left(
m_1^8 m_2^4
  (2\sigma^2-1)^2
          \right)          }{((p_1+l_1)^2-m_1^2+i\varepsilon)((p_1-l_2)^2-m_1^2+i\varepsilon)}\cr
                    \times \frac{1}{((p_1-l_1-l_2-q)^2-m_1^2+i\varepsilon)((p_2-l_1)^2-m_2^2+i\varepsilon)l_1^2 (l_1+q)^2 (l_1+l_2)^2}\,,
\end{multline}
\begin{multline}\label{e:Jseivs}
  J_{SE}^{IV,s}=\begin{gathered} \begin{fmffile}{SEIV}
    \begin{fmfgraph*}(100,100)
\fmfstraight
\fmfleftn{i}{2}
\fmfrightn{o}{2}
\fmf{fermion,label=$p_1$,label.side=left}{i1,vvv1}
\fmf{plain,tension=2}{vvv1,v1}
\fmf{fermion,label=$p_1'$,label.side=left}{vv2,i2}
\fmf{plain,tension=2}{v2,vv2}
\fmf{fermion,label=$p_2'$,label.side=right}{vv3,o2}
\fmf{plain,tension=2}{vv3,v3}
\fmf{fermion,label=$p_2$,label.side=right}{o1,vv4}
\fmf{plain,tension=2}{v4,vv4}
\fmf{dbl_wiggly,tension=0.5}{v4,v1}
\fmf{dbl_wiggly,tension=0.5}{v2,v3}
\fmf{plain,tension=2}{mm1,v2}
\fmf{fermion}{m1,mm1}
\fmf{plain,tension=2}{v1,m1}
\fmf{fermion}{mm3,m3}
\fmf{plain,tension=2}{v4,mm3}
\fmf{plain,tension=2}{v3,m3}
\fmf{phantom,left,tension=0.}{m1,vv2}
\fmf{phantom,left,tension=0.}{vvv1,vv2}
\fmf{phantom,left,tension=0.}{m3,vv3}
\fmf{phantom,left,tension=0.}{m3,vv4}
\fmf{phantom,left,tension=0.}{m3,mm3}
\fmf{dbl_wiggly,left,tension=-0.2}{m1,mm1}
\fmf{phantom,left,tension=0}{vvv1,vv2}
\fmf{phantom,left,tension=0}{vv3,vv4}
\end{fmfgraph*}
\end{fmffile}
\end{gathered}
                 =  \int \frac{d^D l_1 d^D l_2}{(2\pi)^{2D}} \frac{ (16\pi G_N)^3 \hbar^7\left(
m_1^8 m_2^4 (2\sigma^2-1)^2
          \right)         }{((p_1+l_1)^2-m_1^2+i\varepsilon)^2((p_1-l_2)^2-m_1^2+i\varepsilon)}\cr
                   \times \frac{1}{((p_2-l_1)^2-m_2^2+i\varepsilon)l_1^2 (l_1+q)^2 (l_1+l_2)^2}\,,
\end{multline}
The complete classical part of the two-loop amplitude~\eqref{e:M2}  is then finally given by
\begin{multline}\label{e:ResultM2}
\mathcal M_2(\gamma,\underline q^2)\Big|_{\rm classical}=\frac{4(4\pi
  e^{-\gamma_E})^{4-D} \pi G_N^3 m_1^2 m_2^2 }{3
   (4-D)|\underline q|^{2(4-D)}
  \hbar}\Bigg[\frac{3(2\gamma^2-1)^3\mathcal E_{\rm C.M.}^2}{(\gamma^2-1)^2}\cr
+\frac{2im_1 m_2(2\gamma^2-1)}{\pi (4-D)
  (\gamma^2-1)^{\frac{3}{2}}}\left(\frac{1-49\gamma^2+18
    \gamma^4}{5}-\frac{6 \gamma(2 \gamma^2-1)(6
    \gamma^2-7)\arccosh(\gamma)}{\sqrt{\gamma^2-1}}\right)\cr
-\frac{9(2\gamma^2-1)(1-5\gamma^2) \mathcal E_{\rm C.M.}^2}{2(\gamma^2-1)}+\frac{3}{2}(m_1^2+m_2^2)(18\gamma^2-1)-m_1
m_2\gamma(103+2\gamma^2)\cr
+\frac{12 m_1 m_2(3+12
  \gamma^2-4\gamma^4)\arccosh(\gamma)}{\sqrt{\gamma^2-1}} \cr
-\frac{12i m_1 m_2(2\gamma^2-1)^2}{\pi (4-D) \sqrt{\gamma^2-1}}
\frac{1+i\pi {4-D\over2}}{(4(\gamma^2-1))^{4-D\over2}}\left(-{11\over3}+\frac{d}{d\gamma} \left(\frac{(2\gamma^2-1)\arccosh(\gamma)}{\sqrt{\gamma^2-1}} \right)\right)\Bigg]\,.
\end{multline}
In the last line we have used a curiously simplified way of expressing part of the result as a $\gamma$-derivative. Writing it this way is an {\em a posteriori} observation that does not
fall out of our way of computing this part. It should be noted that
there are, perhaps not surprisingly, both real and imaginary parts of
the classical part of the amplitude. In particular, the real and
imaginary parts of the last line of~\eqref{e:ResultM2},
\begin{multline}\label{e:M2RR}
\mathcal M^{\rm RR}_2(\gamma,\underline q^2)\Big|_{\rm classical}  =\frac{4(4\pi
  e^{-\gamma_E})^{4-D} \pi G_N^3 m_1^2 m_2^2 }{3
   (4-D)|\underline q|^{2(4-D)}
  \hbar}\Bigg[\frac{12 m_1 m_2(3+12
  \gamma^2-4\gamma^4)\arccosh(\gamma)}{\sqrt{\gamma^2-1}} \cr
-\frac{12i m_1 m_2(2\gamma^2-1)^2}{\pi (4-D) \sqrt{\gamma^2-1}}
\frac{1+i\pi {4-D\over2}}{(4(\gamma^2-1))^{4-D\over2}}\left(-{11\over3}+\frac{d}{d\gamma} \left(\frac{(2\gamma^2-1)\arccosh(\gamma)}{\sqrt{\gamma^2-1}} \right)\right)\Bigg]\,,  
\end{multline}
are seen to satisfy the relation 
\begin{equation}
  \lim_{D\to4} 2(4-D) \Re\eqref{e:M2RR}= -\lim_{D\to4} (4-D)^2\pi \Im\eqref{e:M2RR},  
\end{equation}
which was first argued on the basis of analyticity and crossing symmetry in~\cite{DiVecchia:2021ndb,DiVecchia:2021bdo}.
The additional real part from~\eqref{e:M2RR} plays a crucial role in the story as will be briefly reviewed below.
\section{The scattering angle from the eikonal formalism}
In quantum field theory the scattering amplitude provides us with the scattering cross section, computed order by order in perturbation theory by means of the Born expansion.
Retaining only the classical information from the quantum mechanical
scattering amplitude we might think that we should be able to compute the actual classical trajectories and not just the (classical) cross section. But in general relativity 
the trajectories of two massive objects scattering off each other will penetrate into regions where the metric will be non-trivial and hence coordinate dependent.
Instead, if we compute the classical scattering angle from Minkowski space at far infinity to Minkowski space at far infinity this is an unambiguous and coordinate-independent
quantity. The whole quantum mechanical thinking in terms of perturbation theory is in fact ideally set up to describe this situation, and we can immediately proceed even when
we retain only the classical information. This is the backbone of the post-Minkowskian expansion which uses Minkowski space to define observables in a manner 
completely analogous to the quantum field theoretic expansion based on the Interaction-picture Hamiltonian and matrix elements evaluated on a basis of free fields.\\[10pt]
While the Hamiltonian framework can be used to compute the scattering angle, it is easier at two-loop order to employ the eikonal formalism.
For this, one again converts the amplitude to $b$-space\footnote{This is
  not the impact parameter $b_J$ orthogonal to the asymptotic momentum
  in the center-of-mass frame. The relation between the two quantities
  is $b_J=b \cos(\chi/2)$~\cite{Parra-Martinez:2020dzs,DiVecchia:2021bdo}.} by performing a Fourier
transform with respect to the momentum transfer,
\begin{equation}\label{e:FT}
  \mathcal M_L(\gamma,b,\hbar)={1\over 4m_1m_2\sqrt{\gamma^2-1}} \int_{\mathbb
    R^{D-2}} {d^{D-2}\vec {\underline q}\over (2\pi)^{D-2}} \mathcal
  M_L(\gamma,\underline q^2,\hbar) e^{i\vec {\underline q}\cdot \vec b}\,.
\end{equation}
The classical eikonal phase $\delta(\gamma,b)$ is defined by an
exponentiation  of the $S$-matrix in $b$-space,
\begin{equation}\label{e:Texp}
  1+i\mathcal T=\left(1+i2\Delta(\gamma,b,\hbar)\right) e^{2i\delta(\gamma,b)\over\hbar}\,.
\end{equation}
where all other terms, order by order in perturbation theory, are kept at linear level and lumped into $\Delta(\gamma,b,\hbar)$ as shown. Contrary to what one might think na\"ively,
this does not imply that this quantity $\Delta(\gamma,b,\hbar)$ simply contains all quantum mechanical bits of the amplitude. Indeed, by expanding the exponent it
is immediately evident that the resulting Laurent expansion in $\hbar$ will combine non-trivially with the terms left at linear level through $\Delta(\gamma,b,\hbar)$.
The fact that the amplitude in $b$-space displays such a neat exponentiation is in fact a manifestation of unitarity of the $S$-matrix \cite{Cristofoli:2020uzm}. In fact, one
can easily understand the exponentiation of the $b$-space amplitude as the eikonal analog of introducing Born subtractions.\\[10pt]
By construction, the eikonal phase $\delta(\gamma,b)$ is independent of $\hbar$ and has a perturbative expansion in $G_N$ through the sum over loops,
\begin{equation}
  \delta(\gamma,b)= \sum_{L\geq0} \delta_L(\gamma,b)\,,
\end{equation}
which is connected  to the Laurent expansion in $\hbar$ of the scattering amplitude
in~\eqref{e:Mhbarexp} in $b$-space, 
\begin{equation}
  1+i\mathcal T = 1+i\sum_{L\geq0}  \mathcal
  M_L(\gamma,b,\hbar)\,.
\end{equation}
Having carefully extracted the classical eikonal contribution at a given loop
order we can then in principle evaluate the scattering angle at this order in perturbation theory by the saddle-point condition
\begin{equation}
  \sin\left(\chi\over2\right)\Big|_{\rm L-PM}  = -\frac{{\cal E}_{\rm CM}}
{m_1m_2\sqrt{\gamma^2-1}}\,{\partial \delta_L(\gamma,b)\over \partial b}\,.
\end{equation}
In practice, also this procedure becomes quite involved once accuracy is increased but it works fine up to two-loop order.
From the results for the amplitude we quoted above we immediately get
\begin{equation}\label{e:delta1}
\delta_0(\gamma,b)=G_N m_1
  m_2\frac{ 2\gamma^2-1 }{ \sqrt{\gamma^2-1}}{(\pi b^2
e^{\gamma_E})^{4-D\over2}\over D-4}+\mathcal O((D-4)^0)\,,
\end{equation}
at tree level. At second post-Minkowskian order we likewise get, straightforwardly from the amplitude,
\begin{equation}
\delta_1(\gamma,b)=G_N^2 (m_1+m_2) m_1 m_2\frac{3 \pi (5\gamma^2-1)}{8 b \sqrt{\gamma^2-1}} (\pi b^2 e^{\gamma_E})^{4-D}+\mathcal O(4-D)\,,
\end{equation}
but at third post-Minkowskian order we need to carefully extract the exponent by taking into account the iterations from lower orders as implied by the
eikonal exponentiation formula  (\ref{e:Texp}). After some algebra, one finds
\begin{multline}\label{e:delta2}
\delta_2(\gamma,b)=\frac{G_N^3 m_1
  m_2 (\pi b^2e^{\gamma_E })^{3(4-D)\over2}}{2b^2
  \sqrt{\gamma^2-1}}\Bigg(\frac{2(12\gamma^4-10\gamma^2+1)\mathcal E_{\rm C.M.}^2}{\gamma^2-1}
\cr
-{4m_1m_2\gamma\over3}(25+14\gamma^2)+\frac{4 m_1 m_2(3+12
  \gamma^2-4\gamma^4)\arccosh(\gamma)}{\sqrt{\gamma^2-1}} \Bigg)\cr
+\frac{2 m_1 m_2(2\gamma^2-1)^2}{\sqrt{\gamma^2-1}}
{1\over
  (4(\gamma^2-1))^{4-D\over2}}\bigg(-\frac{11}{3}+\frac{d}{d\gamma}
\Big(\frac{(2\gamma^2-1)\arccosh(\gamma)}{\sqrt{\gamma^2-1}} \Big)
\bigg)\Bigg) +\mathcal O(4-D)\,.
\end{multline}
This result was first found in ref. \cite{DiVecchia:2020ymx} by means of the relation between real and imaginary parts discussed in the previous section. Here we see
how it actually follows directly from the complete amplitude calculation at third post-Minkowskian order, once all classical parts have been correctly computed.\\[10pt]
Expressing the results in terms of angular momentum $J$ through the relation
\begin{equation}\label{e:Jdef}
  J={m_1m_2\sqrt{\gamma^2-1}\over\mathcal E_{\rm C.M.}}b\cos\left(\chi\over2\right)\,,
\end{equation}
we can derive the scattering angle at the first and second
post-Minkowskian order 
\begin{align}
  \chi_{\rm 1PM}&={2(2\gamma^2-1)\over\sqrt{\gamma^2-1}}\, {G_Nm_1m_2\over J},\cr
\chi_{\rm 2PM}&={3\pi \over
            4}{ m_1+m_2\over
            \mathcal E_{\rm C.M.}} (5\gamma^2-1)\left(
            G_Nm_1m_2\over J\right)^2\,.
\end{align}
To this order they can in fact be related to the scattering of a test particle of mass 
$m_1m_2/(m_1+m_2)$ in a static Schwarzschild background of mass $m_1+m_2$.\\[10pt]
At third post-Minkowksian order the result gets more interesting.
Collecting the contributions from the 3-graviton cuts and the
self-energy diagrams the scattering angle reads
\begin{multline}\label{e:Chi3PM}
{  \chi}_{\rm 3PM}= \frac{2 \left(64 \gamma
    ^6-120 \gamma ^4+60 \gamma ^2-5\right)}{3 \left(\gamma
    ^2-1\right)^{3\over2}}\left(G_Nm_1m_2\over J\right)^3\cr
+{8m_1m_2\sqrt{\gamma^2-1}\over 3\mathcal E_{\rm C.M.}^2}\Bigg(-\gamma(25+14\gamma^2)+\frac{3(3+12
  \gamma^2-4\gamma^4)\arccosh(\gamma)}{\sqrt{\gamma^2-1}}\Bigg)
\left(G_Nm_1m_2\over J\right)^3\cr
+{1\over
  (4(\gamma^2-1))^{4-D\over2}}\left(-\frac{11}{3}+\frac{d}{d\gamma}
\left(\frac{(2\gamma^2-1)\arccosh(\gamma)}{\sqrt{\gamma^2-1}}
\right) \right)\cr
\times  {4m_1m_2(2\gamma^2-1)^2\over
    \mathcal E^2_{\rm C.M.}}\left(G_Mm_1m_2\over J\right)^3 +\mathcal O(4-D)\,.
\end{multline}
At this order  the scattering angle deviates from the geodesic
scattering of a test particle by the contribution in the second
line. 
The third line of~\eqref{e:delta2}. is what can be viewed as radiation-reaction terms
in~\cite{Damour:2020tta,DiVecchia:2021ndb,Herrmann:2021tct,Bjerrum-Bohr:2021din}.  The classical piece of
the two-loop amplitude without gravitational radiation is the sum of all these contributions.  Because of infrared singularities the scattering amplitude has
been dimensionally regularized at intermediate stages but the final classical result is infrared finite in the limit $D \to 4$. If we had computed
the amplitude in a truncated region of integration such as that of the so-called potential region we would have missed these additional classical terms.
In fact, it is not natural at this third post-Minkowskian order to separate the pieces that {\em a posteriori} can be understood as
radiation-reaction terms from the other pieces because all we do is to compute the complete classical contribution from the two-to-two scattering amplitude. 
At fourth post-Minkowskian order the amplitude even contains a divergent piece in four dimensions if the integrations are restricted to the so-called potential 
region~\cite{Bern:2021dqo}. It is only cancelled after the inclusion
of all diagrams and an inclusion of all integration regions that contribute to the classical result~\cite{Bern:2021yeh}.
The radiation-reactions contributions have also been derived using
different amplitude based methods: (1)
High-energy scattering~\cite{DiVecchia:2020ymx,DiVecchia:2021ndb}, (2)
 Linear response to the angular
 momentum~\cite{Damour:2020tta,Veneziano:2022zwh,Manohar:2022dea}, (3) 
     Reverse unitarity and the KMOC\footnote{See, chapter 14 of this review~\cite{KMOCsagex}.}
      formalism~\cite{Herrmann:2021lqe,Kosower:2018adc,
        Mougiakakos:2021ckm,Riva:2021vnj}.\\[10pt]
From the scattering angle, one can reconstruct a classical potential
$\mathcal V_{L+1}(r,p)$ in~\eqref{e:H} that produces it by matching the expression for
the angle from the Hamiltonian
formalism~\cite{Damour:2016gwp,Bjerrum-Bohr:2019kec}. We will return to this below.
\subsection{Maximal supergravity}
Although not of physical interest it is nevertheless illuminating to compare the results of Einstein gravity
with the case of the two-body scattering in  maximal supergravity~\cite{Caron-Huot:2018ape,DiVecchia:2020ymx,Parra-Martinez:2020dzs,Bjerrum-Bohr:2021vuf}.
At first Post-Minkowskian order the eikonal phase of $\mathcal{N}=8$ supergravity is given by 
\begin{equation}
\delta^{\mathcal N=8}_1(\gamma,b)=   m_1m_2G_N \frac{2\gamma^2 }{\sqrt{\gamma^2-1}}
{(b\sqrt\pi)^{4-D}\over D-4}+\mathcal O((D-4)^0)\,.
\end{equation}
This expression is quite similar to the one for Einstein gravity
in~\eqref{e:delta1} with the replacement of $2\gamma^2-1$ by
$2\gamma^2$ in the numerator because of cancellations. The external states are massive half-BPS  states constructed by
Kaluza-Klein reduction~\cite{Caron-Huot:2018ape},  and the 
scattering angle is independant of the relative orientation of the
momenta in the extra dimensions~\cite{DiVecchia:2020ymx,Parra-Martinez:2020dzs,Bjerrum-Bohr:2021vuf}.
Such supersymmetric cancellations lead also to the well-known simplification of
the one-loop amplitude which is given only by the box
integral~\cite{Green:1982sw}. Consequently, a vanishing contribution to the one-loop potential ensues and the
second-order Post-Minkowskian scattering angle equals that of tree level~\cite{Caron-Huot:2018ape} in $D=4$ dimensions. In terms of the eikonal phase
one has the one-loop expression
\begin{equation}
\delta^{\mathcal N=8}_2(\gamma,b)=G_N^2  (m_1+m_2)  m_1 m_2 \frac{2\pi \gamma ^4 
 }{ b \left(\gamma
   ^2-1\right)^{3\over2} } (D-4)+\mathcal O((D-4)^2)\,,
\end{equation}
which indeed vanishes in four dimensions.\\[10pt]
At two-loop order the supersymmetric cancellations make the tensorial
reductions much simpler than those of Einstein gravity, but the basis of master integrals is identical to that of
Einstein gravity. One finds~\cite{Bjerrum-Bohr:2021vuf}:
\begin{multline}
 \delta^{\mathcal N=8}_3(\gamma,b)=\frac{8  G_N^3m_1^2 m_2^2\gamma^4}{ b^2} \Bigg[
 -{\arccosh(\gamma)\over \gamma^2-1}\cr
+\bigg(\frac{1}{4(\gamma^2-1)}
 \bigg)^{4-D\over2} {1\over 2(\gamma^2-1)} {d\over d\gamma}\left(
 2\gamma^2\arccosh(\gamma)\over \sqrt{\gamma^2-1}
  \right) +{\cal O}(D-4) \Bigg]\,.
\end{multline}
The scattering angle at the third post-Minkowskian order 
in maximal supergravity is thus given by 
\begin{multline}
  \label{e:Chi3PMN8}
    \chi_{\rm 3PM}^{\mathcal N=8}=-16\gamma^4\left(\frac{ \gamma^2
    }{3(\gamma^2-1)^{\frac{3}{2}}}+2 
    \frac{m_1 m_2}{\mathcal E^2_{\rm C.M.}}\arccosh(\gamma) \right)\left(G_Nm_1m_2\over
    J\right)^3\cr
  +{16 m_1 m_2\over \mathcal E^2_{\rm C.M.}} {\gamma^4  \over 
   (4 (\gamma^2-1))^{4-D\over2}}\frac{d}{d\gamma}
\left(\frac{2\gamma^2\arccosh(\gamma)}{\sqrt{\gamma^2-1}}
\right) \left(G_N
 m_1m_2\over J\right)^3+\mathcal O(D-4)\,.
\end{multline}
The leading ultra-relativistic limit, $\gamma\gg1$  of 
the third post-Minkowskian scattering angle
in Einstein gravity and maximal supergravity (recall that  the
center-of-mass of mass~\eqref{e:ECM} depend on $\gamma$ as well)   is the same
\begin{equation}
  \lim_{\gamma\to\infty}   \eqref{e:Chi3PM}=  \lim_{\gamma\to\infty}
  \eqref{e:Chi3PMN8}={32\over3} \left(G_N m_1m_2\gamma\over J\right)^3+\mathcal O(\gamma^2)\,,
\end{equation}
as expected by universality of the ultra-relativistic gravitational
scattering~\cite{Bern:2020gjj,DiVecchia:2020ymx,DiVecchia:2021bdo}.
\subsection{Velocity cuts}
\label{sec:velocitycuts}
  Velocity cuts can be viewed as unitarity cuts adapted to the Post-Minkowskian
  expansion. They allow for an efficient extraction of the classical piece of
  order $1/\hbar$  in the Laurent expansion in~\eqref{e:Mhbarexp} of
  the two-body scattering amplitude. The organisation of the
    $\hbar$ expansion of the integrand is closely related to the heavy-mass effective field theory 
    adapted to the binary system  used in~\cite{Brandhuber:2021eyq}.\\[10pt]
The idea of velocity cuts relies on the observation that the
combination of linear propagators
\begin{multline}\label{propid}
\left(\frac{1}{({{}p}_A \cdot{}
 \ell_A\!+\!i\varepsilon)({p}_A{} \cdot \ell_B\!-\!i{}\varepsilon)}\!-\!\frac{1}{({p}_A
 \cdot \ell_B\!+{}\!i\varepsilon)({p}_A \cdot
 \ell_A\!-\!i\varepsilon)}\right)\times\cr
 \left(\frac{1}{({{}p}_B \cdot \ell_A\!-\!i\varepsilon)({p}_B \cdot \ell_C\!+\!i\varepsilon)}\!-{}\!\frac{1}{({p}_B \cdot \ell_C\!-\!i\varepsilon)({p}_B \cdot \ell_A\!+\!i\varepsilon)}\right)\,,
\end{multline}
can be expressed in terms of delta functions
\begin{equation}
\left(\frac{{}\delta({p}_A \cdot {}\ell_A)}{{p}_A \cdot {}\ell_B+i\varepsilon}-\frac{\delta({p}_A \cdot \ell_B)}{{p}_B {}\cdot \ell_A+i\varepsilon}\right)\times\left(\frac{\delta({p{}}_B \cdot {}\ell_C)}{{p}_B \cdot \ell_A+i\varepsilon}-\frac{\delta({p}_B {}\cdot{} \ell_A)}{{p}_B \cdot {} \ell_C+i{}\varepsilon}\right)\,,
\end{equation}
thanks to the identity
\begin{equation}\label{e:PP}
{1\over x+i\varepsilon}-{1\over x-i\varepsilon}= -2i\pi\delta(x)\,.
\end{equation}
This  provides considerable simplifications to reorganise some variety
of propagators in the integrand. We can illustrate this by the case of the
sum of the one-loop box and crossed box diagrams
\begin{align}
  I_\Box&=
\begin{gathered}
\begin{fmffile}{box}
\begin{fmfgraph*}(60,60)
\fmfstraight
\fmfleftn{i}{2}
\fmfrightn{o}{2}
\fmf{fermion,label=$p_1$,label.side=left}{i1,v1}
\fmf{fermion,label=$p_1'$,label.side=left}{v2,i2}
\fmf{fermion,label=$p_2'$,label.side=right}{v3,o2}
\fmf{fermion,label=$p_2$,label.side=right}{o1,v4}
\fmf{fermion}{v1,v2}
\fmf{fermion}{v4,v3}
\fmf{dbl_wiggly}{v4,v1}
\fmf{dbl_wiggly}{v2,v3}
\end{fmfgraph*}
\end{fmffile}
\end{gathered}\qquad +\qquad
\begin{gathered}
\begin{fmffile}{crossbox}
\begin{fmfgraph*}(60,60)
\fmfstraight
\fmfleftn{i}{2}
\fmfrightn{o}{2}
\fmf{fermion,label=$p_1$,label.side=left}{i1,v1}
\fmf{fermion,label=$p_1'$,label.side=left}{v2,i2}
\fmf{fermion,label=$p_2'$,label.side=right}{v3,o2}
\fmf{fermion,label=$p_2$,label.side=right}{o1,v4}
\fmf{fermion,tension=0.01}{v1,v2}
\fmf{fermion,tension=0.01}{v4,v3}
\fmf{dbl_wiggly}{v4,v2}
\fmf{dbl_wiggly}{v1,v3}
\fmf{phantom,tension=0}{v1,v2}
\fmf{phantom,tension=0}{v4,v3}
\end{fmfgraph*}
\end{fmffile}
\end{gathered}\cr
 &=\int \frac{d^D
   \ell}{(2\pi \hbar)^D }\frac{1}{\ell^2 (\ell+q)^2}\left(\frac{1}{(-p_1
   +\ell)^2-m_1^2+i\varepsilon}+\frac{1}{(p_1' +
   \ell)^2-m_1^2+i\varepsilon}\right)\cr
   &\times\left(\frac{1}{(-p_2 + \ell)^2-m_2^2+i\varepsilon}+\frac{1}{(p_2'+ \ell)^2-m_2^2+i\varepsilon}\right)\,.
\end{align}
We scale the  loop variable $\ell=\hbar |\underline q|l$ and set
$\underline q=|\underline q| u_q$,  and shift the external momenta $p_1=\bar p_1+{\hbar \over 2} \underline q$,
$p_1'=\bar p_1'-{\hbar \over 2} \underline q$, $p_2=\bar
p_2-{\hbar\over 2} \underline q$, $p'_2=\bar
p'_2+{\hbar\over 2} \underline q$ to get
\begin{multline}
 I_{\Box}=-\frac{|\vec{\underline q}|^{D-6}}{8 \hbar^2}\int \frac{d^D
   k}{(2\pi)^D} \frac{1}{k^2 (k+u_q)^2}\cr
 \times\left(\frac{1}{\bar{p}_1 \cdot k+\frac{\hbar
       |\vec{\underline q}| u_q \cdot
       k}{2}+i\varepsilon}-\frac{1}{\bar{p_1} \cdot k-\frac{\hbar
       |\vec{\underline q}| u_q \cdot k}{2}-i\varepsilon}\right)\cr
\times\left(\frac{1}{\bar{p}_2 \cdot k-\frac{\hbar |\vec{\underline
         q}| u_q \cdot k}{2}-i\varepsilon}-\frac{1}{\bar{p}_2 \cdot
     k+\frac{\hbar |\vec{\underline q}| u_q \cdot
       k}{2}+i\varepsilon}\right)\,,
\end{multline}
where we have kept the dependence on $|\underline q|$  in the denominators.
We then do a small $\underline
q$ expansion and neglect tadpoles (they are readily shown to not contribute), to get
  \begin{equation}\label{e:Box}
 I_{\Box}=I_{\Box}^{\rm 1-cut}
 +\frac{|\vec{\underline q}|^{D-5}}{16 \hbar}\int \frac{d^D
   l}{(2\pi)^{D-1}}\frac{1}{\ell^2 (\ell+u_q)^2}
 \left(\frac{ {\delta(\bar p_2\cdot l)}}{(\bar p_1\cdot\ell)^2 }+\frac{{\delta(\bar p_1\cdot l})}{(\bar p_2\cdot\ell)^2 }\right) +\mathcal O(|\underline q|^{D-4})\,.
\end{equation}
We have made a repeated use of the identity~\eqref{e:PP} on the linear
propagators from the massive legs,
e.g. $(-p_1+\ell)^2-m_1^2+i\varepsilon= -2\hbar\bar p_1\cdot
l+i\varepsilon+\mathcal O(\hbar^2)$, and defined
\begin{equation}
 I_{\Box}^{\rm 1-cut}=\frac{|\vec{\underline q}|^{D-6}}{4\hbar^2}
 \left(1+\frac{\hbar^2 |\vec{\underline q}|^2 \mathcal E^2_{\rm C.M.}}{4 m_1^2
m_2^2(\gamma^2-1- \frac{\hbar^2 |\vec{\underline q}|^2
   \mathcal E^2_{\rm C.M.}}{4m_1^2m_2^2})}\right)^{\frac{D-5}{2}}\int \frac{d^D k}{(2\pi)^{D-2}}\frac{\delta(\bar{p}_1 \cdot k)\delta(\bar{p}_2 \cdot k)}{k^2 (k+u_q)^2}\,,
\end{equation}
which evaluates to
\begin{equation}
 I_{\Box}^{\rm 1-cut}=\frac{|\vec{\underline q}|^{D-6}}{4 \hbar^2 m_1 m_2 \sqrt{\gamma^2-1}} \left(1- \frac{\hbar^2 |\vec{\underline q}|^2  \mathcal E^2_{\rm C.M.}}{4m_1^2m_2^2(\gamma^2-1)}\right)^{\frac{4-D}{2}}\frac{\Gamma(\frac{D-4}{2})^2 \Gamma(\frac{6-D}{2})}{(4\pi)^{\frac{D-2}{2}} \Gamma(D-4)}\,.
\end{equation}
Because of our conventions, this is the {\em real} part of the box
integral which is related, by unitarity, to the phase-space  integral over two trees. Remarkably, it can be
evaluated in the soft region by an explicit resummation of the
$q$-expansion. The final result has the highly compact form
shown.  The second term
in~\eqref{e:Box} corresponds to the next-to-leading contribution of
eq.~(2.27) of~\cite{Bjerrum-Bohr:2021vuf}  and  leads to a classical contribution from the $D$-dimensional one-loop amplitude.\\[10pt]
The velocity cuts impose delta functions on the massive scalar
legs. With these the delta-function insertions the Feynman integrals that
arise are identical to the ones from the world-line
formalism
in~\cite{Kalin:2020fhe,Mogull:2020sak,Jakobsen:2021smu,Jakobsen:2021zvh,Dlapa:2021npj,Jakobsen:2022fcj}.                 The
correspondence, illustrated  in~\eqref{eq:wl},  between the two formalism arises thanks
identity~\eqref{e:PP} that rewrites linear propagators from the
Feynman $i\varepsilon$ prescription as the retarded Feynman
propagators plus a delta-function.\\[10pt]
This observation can be used to systematize the $\hbar\to0$
expansion of the integrand in terms of a multiple-soft graviton limit
of tree amplitudes. As will be shown below, it permits the re-grouping of
the integrand according to unitarity relations in an exponential
representation of the $S$-matrix. This in turn will identify the precise subtractions from amplitude that leaves behind the classical radial
action~\cite{Damgaard:2021ipf}.
\begin{equation}\label{eq:wl}
\begin{gathered}
    \begin{fmffile}{newother1}
    \begin{fmfgraph*}(100,100)
\fmfleftn{i}{9}
\fmfrightn{o}{9}
\fmf{plain,label.side=left}{i1,v1}
\fmf{plain,label.side=left}{v2,i9}
\fmf{plain,label.side=right}{v3,o9}
\fmf{plain,label.side=right}{o1,v4}
\fmf{plain,tension=0.1}{v1,v2}
\fmf{plain,tension=0.1}{v3,v4}
\fmf{dbl_wiggly}{v2,m1}
\fmf{dbl_wiggly}{v4,m1}
\fmf{dbl_wiggly}{v3,m1}
\fmf{dbl_wiggly}{v1,m1}
\fmf{phantom}{i5,f2}
\fmf{phantom}{f2,f3}
\fmf{dashes,foreground=red}{f3,f4}
\fmf{phantom}{f4,f5}
\fmf{phantom}{f5,f6}
\fmf{dashes,foreground=red}{f6,f7}
\fmf{phantom}{f7,f8}
\fmf{phantom}{f8,o5}
\end{fmfgraph*}
\end{fmffile}
\end{gathered} \longrightarrow
\begin{gathered}
    \begin{fmffile}{newother2}
    \begin{fmfgraph*}(100,100)
\fmfleftn{i}{9}
\fmfrightn{o}{9}
\fmf{phantom}{i1,vv1}
\fmf{phantom}{vv2,i9}
\fmf{phantom}{vv3,o9}
\fmf{phantom}{o1,vv4}
\fmf{dbl_wiggly}{vv1,v1}
\fmf{dbl_wiggly}{v1,vv2}
\fmf{dbl_wiggly}{v1,vv3}
\fmf{dbl_wiggly}{vv4,v1}
\fmfv{decor.shape=circle,decor.filled=hatched, decor.size=3thick}{vv1}
\fmfv{decor.shape=circle,decor.filled=hatched, decor.size=3thick}{vv2}
\fmfv{decor.shape=circle,decor.filled=hatched, decor.size=3thick}{vv3}
\fmfv{decor.shape=circle,decor.filled=hatched, decor.size=3thick}{vv4}
\end{fmfgraph*}
\end{fmffile}
\end{gathered}, \quad
      \begin{gathered}
    \begin{fmffile}{newother7}
    \begin{fmfgraph*}(100,100)
\fmfleftn{i}{9}
\fmfrightn{o}{9}
\fmf{plain,label.side=left}{i1,v1}
\fmf{plain,label.side=left}{v2,i9}
\fmf{plain,label.side=right}{v3,o9}
\fmf{plain,label.side=right}{o1,v3}
\fmf{plain,tension=0.3}{v1,v5}
\fmf{plain,tension=0.3}{v2,v5}
\fmf{dbl_wiggly,tension=0}{v5,m1}
\fmf{dbl_wiggly,tension=0.2}{v2,m1}
\fmf{dbl_wiggly}{v4,m3}
\fmf{dbl_wiggly}{v3,m3}
\fmf{dbl_wiggly,tension=0.2}{v1,m1}
\fmf{dbl_wiggly,tension=1}{m1,m3}
\fmf{phantom}{i6,f2}
\fmf{dashes,foreground=red}{f2,f3}
\fmf{phantom}{f3,f4}
\fmf{phantom}{f4,ff5}
\fmf{phantom}{ff5,fff5}\fmf{phantom}{fff5,f5}
\fmf{phantom}{f5,f6}
\fmf{phantom}{f6,f8}
\fmf{phantom}{f8,o6}
\fmf{phantom}{i4,xf2}
\fmf{dashes,foreground=red}{xf2,xf3}
\fmf{phantom}{xf3,xf4}
\fmf{phantom}{xf4,xff5}
\fmf{phantom}{xff5,xfff5}\fmf{phantom}{xfff5,xf5}
\fmf{phantom}{xf5,xf6}
\fmf{phantom}{xf6,xf8}
\fmf{phantom}{xf8,o4}
\end{fmfgraph*}
\end{fmffile}
\end{gathered} \longrightarrow 
\begin{gathered}
    \begin{fmffile}{newother8}
    \begin{fmfgraph*}(100,100)
\fmfleftn{i}{9}
\fmfrightn{o}{9}
\fmf{phantom}{i1,vv1}
\fmf{phantom}{vv2,i9}
\fmf{phantom}{vv3,i5}
\fmf{phantom}{o5,vv4}
\fmf{phantom}{vv1,vv2}
\fmf{dbl_wiggly,tension=0.2}{vv1,v1}
\fmf{dbl_wiggly,tension=0.2}{v1,vv2}
\fmf{dbl_wiggly,tension=0.2}{v1,vv3}
\fmf{dbl_wiggly}{vv4,v3}
\fmf{dbl_wiggly,tension=1}{v1,v3}
\fmfv{decor.shape=circle,decor.filled=hatched, decor.size=3thick}{vv1}
\fmfv{decor.shape=circle,decor.filled=hatched, decor.size=3thick}{vv2}
\fmfv{decor.shape=circle,decor.filled=hatched, decor.size=3thick}{vv3}
\fmfv{decor.shape=circle,decor.filled=hatched, decor.size=3thick}{vv4}
\end{fmfgraph*}
\end{fmffile}
\end{gathered} 
\end{equation}
\subsection{A multi-soft graviton expansion}
The classical limit $\hbar\to0$ limit of the scattering amplitude in~\eqref{e:Mhbarexp} can
be organized using a multi-soft limit of
the tree-level amplitudes for the emissions of gravitons from a massive
scalar line in the cut integral in~\eqref{e:MLcut}.\\[10pt]
Let us illustrate by considering the tree-level amplitudes ${{}\cal M}^{{}\rm tree}_{
  L+1}(p,\ell_2,\ldots,\ell_{{}L+2},-p')$
and make the substitution 
$\ell_i\to \hbar|\underline{q}| \tilde \ell_i$ with $\hbar\to0$, so that
\begin{equation}
 q= p-p'=-\hbar|\underline q|\sum_{i=2}^{L+2} \tilde \ell_i\,.
\end{equation}
The  tree-level graviton emission amplitude has the universal
behaviour 
\begin{equation}
\lim_{|\vec q|\to0} \mathcal M^{\rm tree}_{ L+1}(p, \hbar|\underline q|\tilde\ell_2,\dots , \hbar|\underline q| \tilde \ell_{L+2},-p') \propto (\hbar|\underline q|)^{-L}\,.
\end{equation}
By repeated use of the identity
\begin{multline}\label{e:propflip}
\frac{1}{(p_1-\ell_{i_2}-\cdots-\ell_{i_j}-q)^2-m^2+i \varepsilon}=
-2\pi i \delta\big((p_1-\ell_{i_2}-\cdots-\ell_{i_j}-q)^2-m^2\big)\cr
+\frac{1}{(p_1-\ell_{i_2}-\cdots-\ell_{i_j}-q)^2-m^2-i \varepsilon}\,,
\end{multline}
we can rewrite the four-point tree-level amplitudes as (setting $\deltabar(x)\equiv-2\pi
i \delta(x)$)
\begin{multline}\label{e:M2plus}
{\mathcal M}^{\rm tree}_2(p_1,\hat\ell_2,\ell_3,-p_1')= {\mathcal M}^{\textrm{tree}(+)}_2(p_1,\hat\ell_2,\ell_3,-p_1')\cr
+\deltabar((p_1+\hat\ell_2)^2-m_1^2)
 {\mathcal M}^{\rm tree}_1(p_1,\hat\ell_2,-p_1-\hat\ell_2)
 {\mathcal M}^{\rm tree}_1(p_1+\hat\ell_2,\ell_3,-p_1')\,,
\end{multline}
and in the five-point case
\begin{multline}
 {\mathcal M}^{\rm tree}_3(p_1,\ell_2,\ell_3,\hat{\ell}_4,-p_1')=
 \deltabar((p_1+\hat{\ell}_4)^2-m_1^2)\deltabar((p_1+\ell_2+\hat{\ell}_4)^2-m_1^2)\cr
 \times
 {\mathcal M}^{\textrm{tree}(+)}_1(p_1,\hat{\ell}_4,-p_1-\hat{\ell}_4){\mathcal M}^{\textrm{tree}(+)}_1(p_1+\hat{\ell}_4,\ell_2,-p_1-\hat{\ell}_4-\ell_2){\mathcal M}^{\textrm{tree}(+)}_1(p_1+\hat{\ell}_4+\ell_2,\ell_3,-p_1')\cr
 +
 \deltabar((p_1+\hat{\ell}_4)^2-m_1^2)\deltabar((p_1+\ell_3+\hat{\ell}_4)^2-m_1^2)
 \cr
 \times
 {\mathcal M}^{\textrm{tree}(+)}_1(p_1,\hat{\ell}_4,-p_1-\hat{\ell}_4){\mathcal M}^{\textrm{tree}(+)}_1(p_1+\hat{\ell}_4,\ell_3,-p_1-\hat{\ell}_4-\ell_3){\mathcal M}^{\textrm{tree}(+)}_1(p_1+\hat{\ell}_4+\ell_3,\ell_2,-p_1')\cr
 +\deltabar((p_1+\ell_2+\hat{\ell}_4)^2-m_1^2)
 {\mathcal M}^{\textrm{tree}(+)}_2(p_1,\ell_2,\hat{\ell}_4,-p_1-\ell_2-\hat{\ell}_4){\mathcal M}^{\textrm{tree}(+)}_1(p_1+\hat{\ell}_4+\ell_2,\ell_3,-p_1')\cr
 +\deltabar((p_1+\ell_3+\hat{\ell}_4)^2-m_1^2){\mathcal M}^{\textrm{tree}(+)}_2(p_1,\ell_3,\hat{\ell}_4,-p_1-\ell_3-\hat{\ell}_4){\mathcal M}^{\textrm{tree}(+)}_1(p_1+\hat{\ell}_4+\ell_3,\ell_2,-p_1')
 \cr
 +
 \deltabar((p_1+\hat{\ell}_4)^2-m_1^2){\mathcal M}^{\textrm{tree}(+)}_1(p_1,\hat{\ell}_4,-p_1-\hat{\ell}_4){\mathcal M}^{\textrm{tree}(+)}_2(p_1+\hat{\ell}_4,\ell_2,\ell_3,-p_1')\cr
 +{\mathcal M}^{\textrm{tree}(+)}_3(p_1,\ell_2,\ell_3,\hat{\ell}_4,-p_1')\,.
\end{multline}
In all generality  we have an expansion organised by powers of
  delta-functions insertions on the massive lines  as given  in eq.~(4.8)
  of~\cite{Brandhuber:2021eyq} and eq.~(4.36) of~\cite{Bjerrum-Bohr:2021wwt} 
\begin{multline}\label{e:MLdelta}
\mathcal M^{\rm tree}_{L+1} \sim (\mathcal M^{\textrm{tree}(+)}_1)^{L+1} \prod_i^L
\delta_i(\ldots) +({\mathcal M}^{\textrm{tree}(+)}_1)^{L-1}
(\mathcal M^{\textrm{tree}(+)}_2) \prod_i^{L-1} \delta_i(\ldots)+\cdots\cr
+\mathcal M^{\textrm{tree}(+)}_1 \mathcal M^{\textrm{tree}(+)}_L \delta(\ldots)+\mathcal M^{\textrm{tree}(+)}_{L+1}\,.
\end{multline}
The tree-level amplitudes $\mathcal
M_{L+1}^{\textrm{tree}(\pm)}(p_1,\ell_2,\dots,\hat\ell_i,\dots,\ell_{L+2},-p_1')$
are defined with the opposite $i\varepsilon$
prescription for the propagators involving a marked graviton leg
$\hat\ell_i$ by using~\eqref{e:propflip}. 
The point of this rewriting is that the amplitudes $\mathcal  M_{L+1}^{\textrm{tree}(\pm)}$ have the multi-soft behaviour 
\begin{equation}\label{e:Mtildescaling}
\lim_{|\vec q|\to0} \mathcal M^{\textrm{tree}(\pm)}_{L+1}(p,
\hbar|\underline q|\tilde\ell_2,\dots , \hbar|\underline
q|\hat{\tilde\ell}_{L+2},-p') \propto (\hbar|\underline q|)^0\,.
\end{equation}
By combining~\eqref{e:Mtildescaling} with the scaling of the
delta function 
\begin{equation}
 \delta{\left((p_1+ \sum \ell_i)^2-m_1^2\right)} = \delta \left(2 \hbar|\underline q| p_1\cdot \sum \tilde\ell_i+\mathcal O(|q|^2)\right)={1\over \hbar|\underline q|} \delta\left(2 p_1\cdot \sum \tilde\ell_i\right)+\mathcal O(|q|^0)\,,
\end{equation}
we deduce that the multi-soft expansion of the tree-level amplitude
$\mathcal M^{\rm tree}_{L+1}$ is organised by the different powers of
delta-function, implementing the velocity cut insertions.
When plugged into the expression for the  integrand of the cut
integral in~\eqref{e:MLcut}, it becomes a sum of contributions
organized as follows
\begin{equation}\label{e:MLcutdelta}
 \mathcal M_{L}^{\rm cut}\sim\sum_{k=0}^{2L}  \hbar^{3L+1} \int {(d^D\ell)^L\over \hbar^{DL}}\frac{
 \left(\delta((p_1+\sum_i \ell_{\alpha_i})^2-m_1^2)\right)^{k} \times (\prod M^{\textrm{tree}(+)}) \times
 (\prod M^{\textrm{tree}(-)\,\dagger})}{(\ell^2)^{L+1}}\,.
\end{equation}
Setting $\ell=\hbar
|\underline q|\hat \ell$ we see that the generic integrals in the multi-graviton cut
behave as
\begin{equation}
 \mathcal M_{L}^{\rm cut}\sim \sum_{k=0}^{2L}  {\hbar^{L-1-k}\over
 |\underline q|^{2+k-(D-2)L}}\,.
\end{equation}
We can have three type of contributions:
\begin{itemize}
\item{ The terms with $k=L$ delta-functions} behave as
 \begin{equation}
     {1\over \hbar |\underline q|^{2-(D-3)L}}\,,
    \end{equation}
    which is of classical order and given by terms with $r=L-2$ in~\eqref{e:Mhbarexp}. Thus in this case,
\begin{equation}\label{e:Mclassical}
 \mathcal M_L(\gamma,|\underline q|^2)\Big|_{\rm{classical}}= {1\over \hbar}
 {\mathcal M_L^{(L-2)}(\gamma,D)\over |\underline q|^{2-(D-3)L}}\,,
\end{equation}
which implies that for computing the classical part of the amplitude,
we can approximate the unitarity delta-function constraint as a velocity
cut $\delta((p+\ell)^2-m^2)\sim \delta(2p\cdot\ell)$, which hugely simplifies 
the integral computation.
\item{ The terms with $k<L$ delta-functions} are of order $\mathcal
 O(\hbar^0)$ and correspond to quantum contributions.
 \item{ The terms with $k>L$ delta-functions} correspond to contributions
  with $-2\leq r\leq L-3$ in the Laurent
  expansion~\eqref{e:Mhbarexp}. These
  contributions are precisely the ones arising from the expansion
  of the exponential representation of the $S$-matrix described in section~\ref{sec:exp}.
  \end{itemize}
  \subsection{Mass expansion of the amplitude}
The  classical $L$-loop amplitude  has the generic mass expansion
\begin{equation}
\mathcal M_L(\gamma,\underline q^2)\Big|_{\rm classical}=
{G_N^{L+1}m_1^2m_2^2\over\hbar  |\underline q|^{2+{(2-D)L\over 2}}}
\sum_{i=0}^{L} c_{L-i+2
,i+2
}(\gamma,D) m_1^{L-i} m_2^{i}\,,
\end{equation}
Using the reorganisation of the $\hbar$ expansion in the cut
multi-loop amplitude from the soft expansion of the tree-level
amplitudes,  the coefficients $c_{i,j}(\gamma,D)$ are
easily identified. This identification of the coefficient is valid for
generic values of the masses $m_1$ and $m_2$.\\[10pt]
For instance, at two-loop order the coefficient $c_{4,2}(\gamma,D)$ is obtained
by evaluating the following graph\\[5pt]
\begin{equation}\label{e:probe}
 m_1^{4}m_2^2 \, c_{3,1}(\gamma,D)\simeq \begin{gathered}
    \begin{fmffile}{probe2}
    \begin{fmfgraph*}(100,100)
\fmfleftn{i}{9}
\fmfrightn{o}{9}
\fmftop{t}
\fmfbottom{b}
\fmfrpolyn{smooth, tension=2,label= $-$}{er}{8}
\fmfv{d.sh=circle,d.f=empty,d.si=.185w}{v2}
\fmfv{d.sh=circle,d.f=empty,d.si=.185w}{v5}
\fmfv{d.sh=circle,d.f=empty,d.si=.185w}{v1}
\fmflabel{$p_1'$}{i9}
\fmflabel{$p_2'$}{o9}
\fmflabel{$p_2$}{o1}
\fmflabel{$p_1 $}{i1}
\fmf{phantom}{i1,xd1,xd3,xd4,xd5,xd7,i9}
\fmfv{label.angle=15,label.dist=0.038w,label=$+$}{xd1}
\fmfv{label.angle=15,label.dist=0.038w,label=$+$}{xd7}
\fmfv{label.angle=0,label.dist=0.038w,label=$+$}{xd4}
\fmf{plain}{i1,v1}  
\fmf{plain}{v2,i9}
\fmf{plain}{er4,o9}
\fmf{plain}{o1,er6}
\fmf{plain,tension=0.3}{v1,v5}
\fmf{plain,tension=0.3}{v2,v5}
\fmf{dbl_wiggly,tension=0}{v5,m1}
\fmf{dbl_wiggly,tension=0.2}{v2,m1}
\fmf{dbl_wiggly,tension=0.2}{v1,m1}
\fmf{plain,tension=10}{m1,er1}
\fmf{phantom}{i6,f2}
\fmf{dashes,foreground=red}{f2,f3,f4}
\fmf{phantom}{f3,f4}
\fmf{phantom}{f4,ff5}
\fmf{phantom}{ff5,fff5}\fmf{phantom}{fff5,f5}
\fmf{phantom}{f5,f6}
\fmf{phantom}{f6,f8}
\fmf{phantom}{f8,o6}
\fmf{phantom}{i4,xf2}
\fmf{dashes,foreground=red}{xf2,xf3,xf4}
\fmf{phantom}{xf3,xf4}
\fmf{phantom}{xf4,xff5}
\fmf{phantom}{xff5,xfff5}\fmf{phantom}{xfff5,xf5}
\fmf{phantom}{xf5,xf6}
\fmf{phantom}{xf6,xf8}
\fmf{phantom}{xf8,o4}
\fmf{dashes,for=black}{b,t}
\end{fmfgraph*}
\end{fmffile}
\end{gathered}
\end{equation}\\[5pt]
where we have imposed two velocity cuts (depicted as red dashed lines)
on the left tree-level factor in the unitarity cut of the amplitude
in~\eqref{e:MLcut}.  The blobs represent the multi-graviton
  tree-level emission amplitudes constructed
  in~\cite{Bjerrum-Bohr:2021wwt} based on the tree construction outlined in ref.~\cite{Bjerrum-Bohr:2019nws,Bjerrum-Bohr:2020syg}.  Similar graphs arise in the heavy-mass effective theory used in~\cite{Brandhuber:2021kpo,Brandhuber:2021eyq,Brandhuber:2021bsf}.  The coefficient $m_1^2m_2^{4}
\,c_{2,4}(\gamma,\underline q^2)$ is
obtained by putting the velocity cuts on the right  tree-level factor
in the  cut of the amplitude
in~\eqref{e:MLcut}. With the obvious generalisation for higher-loop
order coefficients
$c_{2+L,2}(\gamma,D)$ and $c_{2,2+L}(\gamma,D)$~\cite{Brandhuber:2021eyq,Bjerrum-Bohr:2021wwt}.
The coefficient $c_{3,3}(\gamma,D)$ is obtained by the graphs 
\smallskip\\[5pt]
\begin{equation}
  m_1^3m_2^2 c_{3,3}(\gamma,D)\simeq
  \begin{gathered}
     \begin{fmffile}{nexttoprobe} 
    \begin{fmfgraph*}(200,150) 
\fmfleftn{i}{9}
\fmfrightn{o}{9}
\fmftop{t}
\fmfbottom{b}
\fmfrpolyn{phantom,smooth,filled=0}{el}{8}
\fmfrpolyn{phantom,smooth}{ert}{8}
\fmfrpolyn{phantom,smooth}{erb}{8}
\fmflabel{$p_1$}{i1}
\fmflabel{$p_1'$}{i9}
\fmflabel{$p_2'$}{o9}
\fmflabel{$p_2$}{o1}
\fmf{phantom}{i1,xxxd1,xd1,,xxd1,xxd3x,xxd3,xd3,xd4,xd5,xxd5,xxd7,xxd7x,xd7,xxxd7,i9}
\fmfv{label.angle=-25,label.dist=-0.0008w,label=$\!\!+$}{xd1}
\fmfv{label.angle=25,label.dist=-0.0008w,label=$\!\!+$}{xd7}
\fmfv{d.sh=circle,d.f=empty,d.si=.12w}{v5}
\fmfv{d.sh=circle,d.f=empty,d.si=.12w}{mm1}
\fmfv{d.sh=circle,d.f=empty,d.si=.12w}{m1}
\fmfv{d.sh=circle,d.f=empty,d.si=.12w}{v2}
\fmfv{d.sh=circle,d.f=empty,d.si=.12w}{v1}
\fmflabel{$+\,$}{xxxfffff8}
\fmflabel{$\,\_$}{xxfffff8}
\fmflabel{$\,\_$}{xfffff8}
\fmf{plain,label.side=left}{i1,v1}
\fmf{plain,label.side=left}{v2,i9}
\fmf{dbl_wiggly,tension=0}{v1,mm1}
\fmf{dbl_wiggly,tension=0}{v5,mm1}
\fmf{dbl_wiggly,tension=0}{v5,m1}
\fmf{dbl_wiggly,tension=0}{v2,m1}
\fmf{plain,tension=0.3}{mm1,erb8}
\fmf{plain,tension=0.3}{v1,v5}
\fmf{plain,tension=0.3}{v2,v5}
%
\fmf{plain,label.side=right}{el4,v2}
\fmf{plain,label.side=right}{ert3,o9}
\fmf{plain,label.side=right,foreground=black}{ert3,ert7}
\fmf{plain,label.side=right}{ert7,v11}
\fmf{plain,label.side=right,foreground=black}{v11,erb7}
\fmf{plain,label.side=right}{o1,erb7}
\fmf{plain,tension=0.3 }{m1,ert3}
\fmf{dashes,foreground=red}{i7,xxf4}
\fmf{phantom}{xxf4,xxff5}
\fmf{phantom}{xxff5,xxfff5}
\fmf{phantom}{xxfff5,xxff8}
\fmf{phantom}{xxff8,xxffff8}
\fmf{phantom}{xxffff8,xxfffff8}
\fmf{phantom}{xxfffff8,o7}
\fmf{dashes,foreground=red}{i3,xf4}
\fmf{phantom}{xf4,xff5}
\fmf{phantom}{xff5,xfff5}
\fmf{phantom}{xfff5,xff8}
\fmf{phantom}{xff8,xffff8}
\fmf{phantom}{xffff8,xfffff8}
\fmf{phantom}{xfffff8,o3}
\fmf{dashes,foreground=red}{o5,xxxf4}
\fmf{phantom}{xxxf4,xxxff5}
\fmf{phantom}{xxxff5,xxxff8}
\fmf{phantom}{xxxff8,xxxffff8}
\fmf{phantom}{xxxffff8,xxxfffff8}
\fmf{phantom}{xxxfffff8,i5}
\fmf{dashes,for=black}{b,t}
\end{fmfgraph*}
\end{fmffile}\end{gathered}.
\end{equation}\\[5pt]
And so, on, the other monomials in the masses are obtained by 
distributing the velocity cuts on each side of the tree factor in the
multiple graviton cut in~\eqref{e:MLcut}.
\section{An exponential representation of the $S$-matrix}\label{sec:exp}
The approaches presented above do lead to fairly complicated computations. In
the first place, the eikonal exponentiation in~\eqref{e:Texp} is
obtained after a  careful separation, order by order, of the various
terms that go into the exponent and those terms that must remain as
prefactor at the linear level. A
second complication is that  after exponentiation in impact-parameter
space one must apply the inverse transformation and seek from it two
crucial ingredients: (1) the correct identification of the transverse
momentum transfer $\vec {\underline q}$ in the center-of-mass frame and (2) the
correct identification of the scattering angle from the saddle point.
At low orders in the eikonal expansion, this procedure works well but
it hinges on the impact-parameter transformation being able to undo
the convolution product of the momentum-space representation. When
$\underline q^2$-corrections are taken into account (and they do need to be included at higher orders) it is well-known
that this procedure requires amendments. While doable in principle, it becomes increasingly difficult with each new order of accuracy.
This motivates why we must seek alternative pathways. One particularly promising method is rooted in the semi-classical WKB
approximation rather than the eikonal formalism. \\[10pt]
The eikonal formalism is built on the transformation of the scattering amplitude to $b$-space by the Fourier transform discussed above. With care, this can
exponentiated, leaving the rest of the amplitude at linear level. But only what eventually exponentiates contributes to the eikonal saddle-point condition.  It would be
interesting if one could reverse the order in which one takes the $b$-space transform so that one instead considers matrix elements of an exponent from the 
outset. Such seems to be the prescription suggested in refs.~\cite{Bern:2021dqo,Bern:2021yeh}, although it is not expressed in those terms there. If we
could pursue such a strategy of working with the $S$-matrix in an exponential representation {\em and} consider the radical idea of computing matrix elements 
of that operator in the exponent rather than the $S$-matrix itself we might effectively be computing matrix elements of a phase shift operator. By extension,
this should be the WKB-limit of the $S$-matrix, and the phase would then be the radial action 
\begin{equation}
  \mathcal S(r,\varphi; E,J)=  J \varphi +\int p_r(r;E,J ) dr\,,
\end{equation}
as known from the match to the Hamilton-Jacobi equation in that semi-classical limit. If possible, this should provide a most efficient way to compute the classical
scattering angle from amplitudes.\\[10pt]
To this end, let us consider the exponential representation of the $S$-matrix at the operator level that has recently been introduced in~\cite{Damgaard:2021ipf},
\begin{equation}\label{e:Sexp}
  \widehat S= \mathbb I+{i\over\hbar}\widehat T= \exp\left(i \widehat N\over\hbar\right)\,,
\end{equation}
with the completeness relation
\begin{equation}\label{e:complet}
\mathbb I = \sum_{n=0}^{\infty} \frac{1}{n!} \int
\prod_{i=1}^2\frac{d^{D-1}k_i}{(2\pi\hbar)^{D-1}} \frac{1}{2E_{k_{1}}}
\prod_{j=1}^n 
\frac{d^{D-1}\ell_j}{(2\pi\hbar)^{D-1}} \frac{1}{2E_{\ell_{j}}}  
\left| k_1, k_2; \ell_1, \ldots, \ell_n \right\rangle\left\langle k_1, k_2; \ell_1, \ldots, \ell_n \right|\,,
\end{equation}
which includes all the exchange of gravitons for $n\geq1$ entering the
radiation-reaction contributions $\hat N^{\rm rad}$.
With this exponential representation of the $S$-matrix, we
 systematically relate matrix elements of the operator in the
 exponential $\hat N$ to ordinary Born amplitudes minus pieces
 provided by unitarity cuts~\cite{Damgaard:2021ipf}. This is 
 seen by the perturbation expansion
 \begin{align}\label{e:NtoT}
 \hat{N}_0  &=  \hat{T}_0, \qquad
 \hat{N}_0^{\rm rad} = \hat{T}_0^{\rm rad}, \cr
\hat{N}_1  &=  \hat{T}_1 - \frac{i}{2\hbar}\hat{T}_0^2, \qquad
\hat{N}_1^{\rm rad} = \hat{T}_1^{\rm rad} -\frac{i}{2\hbar}(\hat{T}_0\hat{T}_0^{\rm rad} + \hat{T}_0^{\rm rad}\hat{T}_0) ,\cr
\hat{N}_2  &=  \hat{T}_2  - \frac{i}{2\hbar}(\hat{T}_0^{\rm rad})^2 - \frac{i}{2\hbar}(\hat{T}_0\hat{T}_1 + \hat{T}_1\hat{T}_0) - \frac{1}{3 \hbar^2}\hat{T}_0^3\,,
\end{align}
and similarly for higher orders. The simplicity of this method seems
very appealing and suggests that it may be used to streamline
post-Minkowskian amplitudes in gravity by means of a diagrammatic
technique that systematically avoids the evaluation of the cut
diagrams that must be subtracted, but simply discards them at the
integrand level. This decomposition is in correspondence with the
$1/\hbar|\underline q|$ expansion of the scattering amplitude
in~\eqref{e:Mhbarexp}. The scattering matrix
operator $\hat T$ is related to  the scattering amplitude $\mathcal
M_L\propto {1\over\hbar} \langle p_1,p_2|\hat
T_L|p_1',p_2'\rangle$. 
The tree-level matrix element for the two-body scattering 
$\mathcal M_0\propto{1\over \hbar}\langle p_1,p_2|\hat
T_0|p_1',p_2'\rangle$ is  of order $\mathcal{O}(1/\hbar)$.
At one-loop order amplitude decomposes into two pieces
\begin{equation}\label{e:M1dec}
  \mathcal
M_1\propto {1\over \hbar}\langle p_1,p_2|\hat
T_1|p_1',p_2'\rangle\propto {1\over \hbar}\langle p_1,p_2|\hat
N_1|p_1',p_2'\rangle+{i\over 2\hbar^2} \langle p_1,p_2|\hat
T_0^2|p_1',p_2'\rangle\,.
\end{equation}
By unitarity the coefficient of the $\mathcal{O}(1/\hbar^2)$
contribution in the scattering amplitude is  $\langle p_1,p_2|\hat
T_0^2|p_1',p_2'\rangle$, and the matrix element $\langle
p_1,p_2|\hat N_1|p_1',p_2'\rangle$ is given by the classical piece is of
order $\mathcal{O}(1/\hbar)$. Therefore, for the classical two-body
scattering only the matrix elements of $\hat N$ are needed.\\[10pt]
The explicit expression for the full one-loop two-body scattering amplitude
in general relativity becomes~\cite{Damgaard:2021ipf}
\begin{equation}\label{e:M1exp}
\mathcal M_1(|\vec{\underline q}|,\gamma,\hbar)={i\hbar\over2} (16\pi G_N
m_1^2m_2^2(2\gamma^2-1) )^2 I_{\Box}^{\rm 1-cut}+ N_1(|\vec{\underline
  q}|,\gamma)+\mathcal O(\hbar)\,,
\end{equation}
and it hence follows that the one-loop contribution to $N$ is
\begin{multline}\label{e:N1}
N_1(|\vec{\underline q}|,\gamma)= \frac{3 \pi^2  G_N^2 m_1^2
  m_2^2(m_1+m_2) (5\gamma^2-1)(4\pi
  e^{-\gamma_E})^{4-D\over2}}{|\vec{\underline q}|^{5-D}} \cr
-\frac{8G_N^2m_1^2 m_2^2(4\pi e^{-\gamma_E})^{4-D\over2}\hbar}{(4-D) |\vec{\underline q}|^{4-D}} \Big(\frac{2(2\gamma^2-1)(7-6\gamma^2)\arccosh(\gamma)}{(\gamma^2-1)^{\frac{3}{2}}}+\frac{1 - 49 \gamma^2 + 18 \gamma^4}{15(\gamma^2-1)}\Big)\cr+\mathcal O(|\vec{\underline q}|^{5-D})\,.
\end{multline}
We now quote from~\cite{Damgaard:2021ipf} the two-loop result in Einstein gravity
\begin{multline}
\mathcal M_2(|\vec{\underline
  q}|,\gamma)={\hbar\over6}(16 \pi G_N m_1^2 m_2^2
  (2\gamma^2-1))^3  I_{\Box \Box}^{\rm 2-cut}\cr
+\frac{12i \pi^2 G_N^3
  (m_1+m_2) m_1^3 m_2^3(2\gamma^2-1)(1-5\gamma^2)(4\pi
  e^{-\gamma_E})^{4-D}}{(4-D)
  \sqrt{\gamma^2-1}|\vec{\underline q}|^{9-2D}}\cr
+\frac{4 \pi
  G_N^3(4\pi e^{-\gamma_E})^{4-D} m_1^2 m_2^2\hbar}{ (4-D)
  |\vec{\underline q}|^{8-2D}}\Bigg(\frac{2im_1
  m_2(2\gamma^2-1)}{\pi (4-D)
  (\gamma^2-1)^{\frac{3}{2}}}\Big(\frac{1-49\gamma^2+18
  \gamma^4}{15}-\frac{2\gamma(7-20 \gamma^2+12
  \gamma^4)\arccosh(\gamma)}{\sqrt{\gamma^2-1}}\Big)\cr
+\frac{\mathcal E^2_{\rm C.M.}\left(64 \gamma ^6-120 \gamma ^4+60 \gamma
   ^2-5\right)}{3\left(\gamma ^2-1\right)^2}-{4 m_1 m_2 \gamma  \left(14 \gamma
   ^2+25\right)\over3}
+\frac{4 m_1 m_2(3+12
  \gamma^2-4\gamma^4)\arccosh(\gamma)}{\sqrt{\gamma^2-1}} \cr-\frac{4i
  m_1 m_2(2\gamma^2-1)^2}{\pi (4-D) \sqrt{\gamma^2-1}} {\frac{1+i\pi
  {4-D\over2}}{(4(\gamma^2-1))^{4-D\over2}}}\bigg(-\frac{11}{3}+\frac{d}{d\gamma}
\Big(\frac{(2\gamma^2-1)\arccosh(\gamma)}{\sqrt{\gamma^2-1}} \Big)
\bigg)\Bigg)+\mathcal O(\hbar)\,,
\end{multline}
where we have introduced the two-cut integral
\begin{multline}
I_{\Box\Box}^{\rm 2-cut} = \left(1-\frac{(11-2D)|\vec{ q}|^2 \mathcal E^2_{\rm C.M.}}{12
  m_1^2 m_2^2 (\gamma^2-1)}\right){1\over16\hbar^{2D-7}}\int \frac{d^D
  l_1 d^D l_2}{(2\pi)^{2D-4}} \frac{\delta(\bar{p}_1 \cdot
  l_1)\delta(\bar{p}_1 \cdot l_2)\delta(\bar{p}_2 \cdot
  l_1)\delta(\bar{p}_2 \cdot l_2)}{l_1^2 l_2^2 (l_1+l_2-q)^2}\cr+\mathcal
  O(|\vec{q}|^{1-4\epsilon})
\end{multline}
which is evaluated to
\begin{multline}
I_{\Box\Box}^{\rm 2-cut} =
-\left(1-\frac{(11-2D)\hbar^2|\vec{\underline q}|^2 \mathcal E^2_{\rm C.M.}}{12
    m_1^2 m_2^2 (\gamma^2-1)}\right){1\over 16\hbar^3 |\vec{\underline
    q}|^{10-2D}}\cr
\times {1\over m_1^2
  m_2^2(\gamma^2-1)-\frac{\hbar^2|\vec{\underline q}|^2 \mathcal E^2_{\rm C.M.}}{4}}\frac{\Gamma\left(D-4\over2\right)^3 \Gamma(5-D)}{(4 \pi)^{D-2} \Gamma\left(3(D-4)\over2\right)}+\mathcal O(|\vec{q}|^{D-7})\,.
\end{multline}
Subtractions of tree and one-loop terms as dictated by
eq.~\eqref{e:NtoT} lead to the following two-loop contribution to $N$
\begin{multline}
N_2(|\vec{\underline q}|,\gamma)=\frac{4 \pi G_N^3(4\pi
  e^{-\gamma_E})^{4-D} m_1^2 m_2^2 }{(4-D) |\vec{\underline q}|^{8-2D}}\Bigg(\frac{\mathcal E^2_{\rm C.M.}\left(64 \gamma ^6-120 \gamma ^4+60 \gamma
   ^2-5\right)}{3\left(\gamma ^2-1\right)^2}\cr-{4\over3} m_1 m_2 \gamma  \left(14 \gamma
   ^2+25\right)+\frac{4 m_1 m_2(3+12
   \gamma^2-4\gamma^4)\arccosh(\gamma)}{\sqrt{\gamma^2-1}}
 \\+\frac{2m_1 m_2(2\gamma^2-1)^2}{
   \sqrt{\gamma^2-1}}\bigg(-\frac{11}{3}+\frac{d}{d\gamma}
 \Big(\frac{(2\gamma^2-1)\arccosh(\gamma)}{\sqrt{\gamma^2-1}} \Big)
 \bigg)\Bigg)+\mathcal O(\hbar)\,.
\end{multline}
We can now collect all contributions to $N$ up to two-loop order,  keeping
only the leading terms in $D-4$, and performing the Fourier
transform to $b$-space. Expressing  everything in 
terms of angular momentum $J$ in~\eqref{e:Jdef}, we get
\begin{multline}
\bar{N}(J,\gamma)=\frac{G_Nm_1 m_2
  (2\gamma^2-1)\Gamma\left(D-4\over2\right)}{\sqrt{\gamma^2-1}}
J^{4-D}+\frac{3 \pi G_N^2 m_1^2 m_2^2(m_1+m_2) (5\gamma^2-1)}{4
  \mathcal E_{\rm C.M.}}\frac{1}{J}\cr
+\frac{G_N^3 m_1^3 m_2^3\sqrt{\gamma^2-1}
}{  \mathcal E_{\rm C.M.}^2}\Bigg(\frac{\mathcal E^2_{\rm C.M.} \left(64 \gamma ^6-120 \gamma ^4+60 \gamma
   ^2-5\right)}{3\left(\gamma ^2-1\right)^2}\cr-{4\over3} m_1 m_2 \gamma  \left(14 \gamma
   ^2+25\right)+\frac{4 m_1 m_2(3+12
   \gamma^2-4\gamma^4)\arccosh(\gamma)}{\sqrt{\gamma^2-1}} \\+\frac{2
   m_1 m_2(2\gamma^2-1)^2}{\sqrt{\gamma^2-1}}
 \bigg(-\frac{11}{3}+\frac{d}{d\gamma}
 \Big(\frac{(2\gamma^2-1)\arccosh(\gamma)}{\sqrt{\gamma^2-1}} \Big)
 \bigg)\Bigg)\frac{1}{J^2}+\mathcal O(\hbar)\,.
\end{multline}
Taking this to be the interacting part of the radial action to third post-Minkowskian order, we obtain the scattering angle
\begin{multline}
\chi=-\frac{\partial}{\partial
  J}\lim_{\hbar\to0}\bar{N}(J,\gamma)=\frac{2
  (2\gamma^2-1)}{\sqrt{\gamma^2-1}} \frac{G_Nm_1m_2}{J}+\frac{3 \pi
  (m_1+m_2) (5\gamma^2-1)}{4 \mathcal E_{\rm C.M.}}\left(G_Nm_1m_2\over J\right)^2\\+2\sqrt{\gamma^2-1} \Bigg(\frac{ \left(64 \gamma ^6-120 \gamma ^4+60 \gamma
   ^2-5\right)}{3\left(\gamma ^2-1\right)^2}\cr-{4m_1m_2\over3 \mathcal E^2_{\rm C.M.}} \gamma  \left(14 \gamma
   ^2+25\right)+\frac{4 m_1 m_2(3+12
   \gamma^2-4\gamma^4)\arccosh(\gamma)}{\mathcal E^2_{\rm C.M.}\sqrt{\gamma^2-1}} \\+\frac{2
   m_1 m_2(2\gamma^2-1)^2}{\mathcal E^2_{\rm C.M.}\sqrt{\gamma^2-1}}
 \bigg(-\frac{11}{3}+\frac{d}{d\gamma}
 \Big(\frac{(2\gamma^2-1)\arccosh(\gamma)}{\sqrt{\gamma^2-1}} \Big)
 \bigg)\Bigg)\left(G_N m_1m_2\over J\right)^3\,,
\end{multline}
which agrees with the literature~\cite{Bern:2019nnu,Antonelli:2019ytb} after inclusion of the crucial 
radiation reaction terms~\cite{DiVecchia:2020ymx,Damour:2020tta,Bjerrum-Bohr:2021din}.
\section{A Post-Minkowskian Effective One-Body formalism}
One route to connect the scattering regime to the bound-state regime is based on the
Effective One-Body (EOB)
formalism~\cite{Buonanno:1998gg,Buonanno:2000ef}, suitably adapted from
post-Newtonian to post-Minkowskian formulations~\cite{Damour:2016gwp,Damour:2017zjx, Damour:2019lcq,Damgaard:2021rnk}.
One important lesson from the scattering amplitude approach to
gravitational scattering in general relativity is that at least up to
and including third post-Minkowskian order, there exists, in isotropic
coordinates, a very simple relationship between center-of-mass
three-momentum $p$ and the effective classical potential $V_{\rm eff} (r, p)$ of the form
\begin{equation}
p^2= p_\infty^2-\mathcal V_{\rm eff}(r,E);  \qquad 
\mathcal V_{\rm eff}(r,E)= -\sum_{n\geq1}f_n \, \left(G_N (m_1+m_2)\over r\right)^n\,,
\end{equation}
where the coefficients $f_n$ are directly extracted from the
scattering angle~\cite{Bjerrum-Bohr:2019kec}
\begin{equation}\label{e:chidef}
  {\chi\over2}= \sum_{k\geq1} {b\over k!} \int_0^\infty du \left(d\over
    du^2\right)^k \left[{1\over u^2+b^2}\left( \mathcal V_{\rm eff}\left(\sqrt{u^2+b^2}\right) (u^2+b^2)\over \gamma^2-1\right)^k  \right]\,.
\end{equation}
Since $p^2=p_r^2+J^2/r^2$ where $J$ is the angular momentum we have
\begin{equation}\label{e:chiAmp}
\frac{\chi}{2} = -\int_{\hat{r}_{m}}^{\infty} dr \frac{\partial
  p_r}{\partial J}  - \frac{\pi}{2}= b \int_{\hat{r}_{m}}^{\infty}
\frac{dr}{r^2} \frac{1}{\sqrt{1 - \frac{b^2}{r^2}- \frac{\mathcal
        V_{\rm eff}(r,E)}{p_{\infty}^2}}}  - \frac{\pi}{2}\,,
\end{equation}
where $\hat r_m$ is solution to $1 - \frac{b^2}{r^2}- \frac{\mathcal
        V_{\rm eff}(r,E)}{p_{\infty}^2}=0$.
Considering a general parametrization of the effective metric $g_{\mu\nu}^{\rm
  eff}$ in isotropic coordinates
\begin{equation}
ds_{\rm eff}^2 = A(r)dt^2 - B(r)\left(dr^2 +r^2(d\theta^2+\sin^2\theta d\varphi^2)\right)\,,
\end{equation}
and the principal function
$\mathcal S= {\cal E}_{\rm eff}t + J_{\rm eff}\varphi + W(r)$
of the associated Hamilton-Jacobi equation
$
g^{\alpha\beta}_{\rm eff}\partial_{\alpha}\mathcal
S\partial_{\beta}\mathcal S = \mu^2
$, the scattering angle is given by
\begin{equation}\label{e:chiEOB}
\frac{\chi}{2} = J_{\rm eff}\int_{r_m}^\infty\frac{dr}{r^2}\frac{1}{\sqrt{{\frac{B(r)}{A(r)}{\cal E}_{\rm eff}^2 - \frac{J_{\rm eff}^2}{r^2} - B(r)\mu^2}}}-{\pi\over2}\,.
\end{equation}
In order to relate the scattering amplitude data in $\mathcal V_{\rm
  eff}(r,E)$ to the effective metric, energy and angular momentum, we reconsider the
maps on which the EOB formalism is based on~\cite{Buonanno:1998gg,
  Buonanno:2000ef,Damour:2012mv}.
We have %
  \begin{itemize}
  \item The {\em energy map}
\begin{equation}
   E=(m_1+m_2)\sqrt{1 + 2 {m_1m_2\over (m_1+m_2)^2}\left(\frac{m_1+m_2}{m_1m_2} {\cal{E}}_{\rm eff}-
        1\right)} 
 \end{equation}
  \item The  {\em momentum map}
\begin{equation}
p_{\infty}^2 ={(E^2-(m_1+m_2)^2)(E^2-(m_1-m_2)^2)\over 4E^2}, \qquad \frac{ p_{\rm eff}}{\mu} ~=~ \frac{p_\infty E}{m_1m_2}
\end{equation}
\item An  {\em angular momentum map}  
\begin{equation}
b ~=~ \frac{J}{p_{\infty}} ~=~ \frac{J_{\rm eff}}{p_{\rm eff}}
\implies J_{\rm eff} ~=~ J{\frac{p_{\rm eff}}{p_{\infty}}}=J\,{E\over M}
\end{equation}
 This map differs from the one used in~\cite{Buonanno:1998gg,
   Buonanno:2000ef,Damour:2012mv}.
 With this relation we choose to keep fixed the impact parameter
   $b$, whereas in~\cite{Buonanno:1998gg}  the angular momentum is kept
   fixed. The possibility of fixing $b$ instead of $J$ has been mentioned in ref.~\cite{Vines:2017hyw } but not pursued there.
\end{itemize}
One can now identify the two expressions in~\eqref{e:chiAmp}
and~\eqref{e:chiEOB} if one equates the expressions under the
square root. This leads to a relation between
the metric coefficients and the effective potential
\begin{equation}
1 - \frac{\mathcal V_{\rm eff}(r,E)}{p_{\infty}^2}={B(r)\over
  \gamma^2-1}\left( \frac{\gamma^2}{A(r)} - 1\right)\,.
\end{equation}
In order to fix the parametrisation ambiguity we  parameterise the
metric coefficient using the {\em Ansatz} 
\begin{equation}
  A(r)=\left(1-{h(r)}\over 1+{h(r)}\right)^2; \qquad B(r)=\left(1+{h(r)}\right)^4\,,
\end{equation}
to get 
\begin{multline}
\left(h(r)+{\gamma-1\over\gamma+1}\right)\left(h(r)+{\gamma+1\over\gamma-1}\right) (1+h(r))^4
\cr
=(1-h(r))^2\left(1+{E^2\over (\gamma^2-1)M^2}{\mathcal V_{\rm eff}(r,E)\over \nu^2M^2}\right)\,.
\end{multline}
Using the known perturbative expansion of the effective potential
$\mathcal V_{\rm eff}=-\sum_{n\geq1} f_n (G_N M/r)^n$  one can solve
for the metric coefficients $h_n$ at each order in $(G M/r)^n$.
The resulting metric deviates from the Schwarzschild solution and
there is no need to introduce the non-metric terms as in the traditional EOB
approach, as they are reabsorbed in the metric coefficients.\\[10pt]
  By computing the two-body scattering in perturbation one derives a
 Lorentz invariant expression valid in all regimes of relative velocity
 between the two interacting massive bodies.
Importantly, the relation between the scattering amplitude and the
Effective-One-Body Effective potential in~\eqref{e:chidef} is valid in
any space-time dimension and
applies to gravity in higher
dimensions~\cite{KoemansCollado:2019ggb,Cristofoli:2020uzm}. 
\section{Summary} \label{sec:summary}
The amplitude approach to the binary problem of general relativity is a highly promising new avenue for computations in the post-Minkowskian approximation. From the scattering
regime, one can infer the dynamics at arbitrarily high energies. From this also the bound pseudo-elliptic regime can be reached and one can hence, eventually, generate waveforms 
that go beyond the Post-Newtonian approximation. While the latter is motivated by the virial theorem, having access to an effective Hamiltonian that is unrestricted in terms of
kinetic energy promises to have definite phenomenological value. Interestingly, the scattering amplitude viewpoint on gravity has also led to a renewed understanding of how 
classical physics can be extracted from quantum mechanical $S$-matrix elements. This interplay is bound to continue in the coming years.\\[10pt]
The general framework presented in this review shows how the Post-Minkowskian expansion from scattering amplitudes can be used to gain complete control of the
two-body scattering in all regimes of energy. As explained, the gravitational radiation field leads to a back-reaction on the participating massive bodies which plays a crucial
role in obtaining correct scattering dynamics already at two-loop order. This is all automatically contained in the ordinary $S$-matrix element of the gravitational two-body
scattering to that order. The radiation reaction to this order thus need not be added separately but is an essential ingredient of the amplitude calculation, correctly taking
into account all contributing diagrams and all loop-integration regimes that lead to classical terms in the amplitude. From this viewpoint the separation of radiation-reaction
terms at two-loop order is not natural since all terms generating classical contributions are already correctly included in the standard Feynman diagram expansion.\\[10pt]
One major technical issue is the extraction of the
classical part of the amplitude (all terms of order  $1/\hbar$ in the Laurent
expansion~\eqref{e:Mhbarexp}).  By combining unitarity and the concept of velocity cuts
introduced in~\cite{Bjerrum-Bohr:2021din}, we can identified exactly
those elements of integrands that lead to classical physics upon
integration. Our approach uses an organisation of the integrand of
the multi-loop amplitude with unitarity cuts on the massive
scalar propagator lines together with detailed knowledge of 
the correspondence between the multi-soft graviton
expansion and $\hbar\to0$ limit classical integrand matching 
the exponential representation of the
$S$-matrix of~\cite{Damgaard:2021ipf}.
In the classical limit, this
approach systematically relates the classical part of the scattering
amplitude to the matrix elements of the operator $\hat N$ of the exponential representation of the $S$-matrix,
without having to actually having to perform the subtractions that are determined by unitarity in a simple manner. Considering the rapid pace at which progress has
taken place we can expect much clarification of the optimal path for amplitude computations for gravity in the coming years.
\section*{Acknowledgments}
This work  was supported  by the European Union's Horizon 2020 research and innovation programme under the Marie Sk\l{}odowska-Curie grant agreement No.~764850 {\it ``\href{https://sagex.org}{SAGEX}''}. The research of P.V. has received funding from the ANR grant ``Amplitude'' ANR-17-
CE31-0001-01, and the ANR grant ``SMAGP'' ANR-20-CE40-0026-01. The work of P.H.D. was supported in part by DFF grant 0135-00089A. The work of N.E.J.B.-B. was supported in part by the Carlsberg Foundation as well as DFF grant 1026-00077B.\\[15pt]
\appendix

\end{document}